\pdfoutput=1
\documentclass[12pt,preprint]{aastex}
\usepackage{graphicx}
\usepackage{natbib}
\usepackage{aas_macros}
\usepackage[section] {placeins}
\usepackage{subfigure}
\usepackage{float}
\usepackage{color}

\def\msun{{\rm\,M_\odot}}

\def\msun{{\rm\,M_\odot}} 

\def\zsun{{\rm\,Z_\odot}}
\newcommand{\etal}{et al.\ }
\newcommand{\kms}{\, {\rm km\, s}^{-1}}

\newcommand{\lya}{Ly$\alpha$ }

\def\h2{${\rm\,H_2}$}

\makeatletter 

\def\etal   {{et~al.}\ }

\def\kms{{\rm\,km/s}}
\def\msun{{\rm\,M_\odot}}

\def\vol#1  {{{#1}{\rm,}\ }}
\def\lya{{\rm Ly}\alpha}

\def\etal{et al.\ }

\newcount\refno
\newcount\rfno
\def\eq{$^{\the\refno\ }$\advance\refno by 1}
\def\ad{\advance\rfno by 1}

\def\clock{\count0=\time \divide\count0 by 60
     \count1=\count0 \multiply\count1 by -60 \advance\count1 by \time
     \number\count0:\ifnum\count1<10{0\number\count1}\else\number\count1\fi}

\def\myputfigure#1#2#3#4#5%
{\hskip0.03\textwidth\vskip#5pt
\makebox[0pt]{\hskip#2in
\includegraphics[width=#3\textwidth]{#1}}\vskip#4pt\hfill}

\newcount\refno
\newcount\rfno
\def\eq{$^{\the\refno\ }$\advance\refno by 1}
\def\ad{\advance\rfno by 1}

\definecolor{burntorange}{rgb}{1,0.4,0.2}

\begin{document}

\title{Coincidences between OVI and OVII Lines:
Insights from High Resolution Simulations of the Warm-Hot Intergalactic Medium}
 
\author{
Renyue Cen$^{1}$
} 
 
\footnotetext[1]{Princeton University Observatory, Princeton, NJ 08544;
 cen@astro.princeton.edu}


\begin{abstract} 

With high resolution ($0.46h^{-1}$kpc), large-scale, adaptive mesh-refinement Eulerian cosmological hydrodynamic simulations 
we compute properties of O~VI and O~VII absorbers from the warm-hot intergalactic medium (WHIM) at $z=0$.
Our new simulations are in broad agreement with previous simulations with $\sim 40\%$ of the intergalactic medium being in the WHIM.
Our simulations are in agreement with observed properties of O~VI absorbers
with respect to the line incidence rate and Doppler width-column density relation. 
It is found that the amount of gas in the WHIM below and above $10^6$K is roughly equal.
Strong O~VI absorbers are found to be predominantly collisionally ionized.
It is found that $(61\%, 57\%, 39\%)$ of O~VI absorbers of
$\log{\rm N(OVI)~cm^2}=(12.5-13,13-14,>14)$ have $T<10^5$K.
Cross correlations between galaxies and strong 
[${\rm N(OVI)}>10^{14}$cm$^{-2}$] O~VI absorbers on $\sim 100-300$kpc scales 
are suggested as a potential differentiator between collisional ionization and photoionization models. 
Quantitative prediction is made for the presence of broad and shallow O~VI lines
that are largely missed by current observations but will be detectable by COS observations.
The reported $3\sigma$ upper limit on the mean column density of coincidental O~VII lines
at the location of detected O~VI lines by Yao \etal is above our predicted value by a factor of $2.5-4$.
The claimed observational detection of O~VII lines by Nicastro et al, if true, is $2\sigma$ above what our simulations predict.

\end{abstract}


\section{Introduction}

Physical understanding of the thermodynamic evolution of the intergalactic medium (IGM) 
has been substantially improved  with the aid of {\it ab initio}
cosmological hydrodynamic simulations.
One of the most robust predictions is that $40-50\%$ of all baryons in the present universe 
is in the WHIM of temperature $10^5-10^7$K and overdensity $10-300$ \citep[e.g.,][]{1999Cen, 2001Dave}.
The predicted WHIM provides an attractive solution to the long standing missing baryons problem
\citep[][]{1992Persic, 1998Fukugita}.
Let us first clarify the nomenclature of several related gas phases.
The intra-group and intra-cluster medium (ICM) is defined to be gas
within these virialized regions (i.e., overdensity $>100$). 
The high density portion (overdensity $\ge 500$) of the ICM 
has traditionally been detected in X-ray emission;
thermal Sunyaev-Zeldovich effect and more sensitive X-ray measurements can now probe 
ICM to about the virial radius.
The circumgalactic medium (CGM) is usually defined to be gas that embeds the stellar components in galactic halos
and may be made up of gases of a wide range of temperatures ($10^4-10^7$K) and densities.
It is likely, at least for large galaxies,
that a significant fraction of the CGM falls into the same temperature range of the WHIM.
Of particular interest is some of the CGM that has been heated up 
by star formation feedback shocks to the WHIM temperature range \citep[e.g.,][]{2006Cen, 2011Cen}.
In the present analysis we define WHIM as gas of temperature $10^5-10^7$K 
with no density limits. 
Most of the WHIM gas is truly intergalactic with overdensity $<100$ (see Figure 7)
and mostly easily probed in absorption.

The reality of the WHIM, at least its low temperature ($T\le 10^6$K) portion,
has now been fairly convincingly
confirmed by a number of observations in the FUV portion of QSO spectra from HST and FUSE,
through the O~VI $\lambda \lambda$1032, 1038 absorption lines
that peak at $T\sim 3\times 10^5$K when collisionally ionized 
\citep[e.g.,][]{2000Tripp, 2000bTripp, 2000Oegerle, 2002Savage, 2004Prochaska,
2004Sembach, 2005Danforth, 2006Danforth, 2008Danforth, 2008Tripp, 2008bThom, 2008Thom, 2008Cooksey}
and Ne VIII $\lambda \lambda$770, 780 absorption lines that peak at $T\sim 7\times 10^5$K in collisional
ionization equilibrium \citep[][]{2005Savage, 2006Savage, 2009Narayanan, 2011Narayanan, 2011Tripp}
as well as broad $\lya$ absorption lines (BLAs) 
\citep[][]{2010Danforth, 2011bSavage, 2011Savage}.
In agreement with simulations, the part of WHIM detected in O~VI absorption 
is estimated to constitute about 20-30\% of total WHIM.
The detection of Ne VIII lines along at least some of the sight lines with O~VI detection
provides unambiguous evidence for the WHIM origin,
instead of lower temperature, photoionized gas,
under physically 
plausible and observationally constrained situations. 

X-ray observations performed to search for X-ray absorption of 
the higher temperature portion ($T\ge 10^6$K) of the WHIM
associated with known massive clusters have also been successful.
An XMM-Newton RGS spectrum of quasar LBQS 1228+1116 revealed a feature at the 
Virgo redshifted position of O~VIII Ly$\alpha$ at the $95\%$ confidence level
\citep[][]{2004Fujimoto}.
Using XMM-Newton RGS observations of an AGN behind the Coma Cluster, 
the Seyfert 1 X Comae,
\citet[][]{2007Takei} claimed to have detected WHIM associated with the Coma cluster. 
Through the Sculptor Wall 
\citet[][]{2009Buote} and \citet[][]{2010Fang}
have detected WHIM O~VII absorption at a column greater than $10^{16}$cm$^{-2}$.
There is evidence of detection in soft X-ray emission along the filament connecting
clusters A222 and A223 at $z=0.21$ that may be associated with the dense and hot portion of the WHIM
\citep[][]{2008Werner}.

However, the search for X-ray absorption of WHIM along random lines of sight
turns out to be elusive.
Early pioneering observations
\citep[][S5 0836+710, PKS 2149-306, PKS 2155-304]{2001Fang, 2002Fang}
gave the first O~VII detection (O~VIII for PKS 2155-304), which has not been convincingly confirmed
subsequently \citep[][]{2004Cagnoni, 2007bWilliams, 2007Fang}.
\citet[][]{2003Mathur} performed a dedicated deep observation 
(470 ks) with the Chandra LETGS of the quasar H 1821+643, 
which has several confirmed intervening O V I absorbers, but found no significant ($>>2\sigma$) X-ray 
absorption lines at the redshifts of the O V I systems. 
\citet[][]{2005Nicastro, 2005bNicastro} embarked on a 
campaign to observe Mrk 421 during its periodic X-ray outbursts
with the Chandra LETGS with a total of more than 7 million continuum counts
and presented evidence for the detection of two intervening absorption systems at $z=0.011$ and $z=0.027$.
But the spectrum of the same source observed with the 
XMM-Newton RGS did not show these absorption lines 
\citep[][]{2007Rasmussen}, despite higher signal-to-noise and comparable spectral resolution.
\citet[][]{2006Kaastra} and \citet[][]{2012Yao} 
reanalyzed the Chandra LETGS data
and were in agreement with \citet[][]{2007Rasmussen}.
The detection of O~VII lines may also be at odds with recent BLA measurements
\citep[][]{2011Danforth}, under simplistic assumptions about the nature of the absorbing medium.
However, it has been argued that the reported XMM-Newton upper limits and the Chandra measurements 
may be consistent with one another, when taking into consideration 
certain instrumental characteristics of the XMM-Newton GRS
\citep[][]{2006Williams}. 
Moreover, an analysis of the two candidate X-ray absorbers at $z=0.011$ and $z=0.027$
yields intriguing evidence of two large-scale filaments at the respective redshifts,
one of which has only $5-10\%$ probability of occurring by chance
\citep[][]{2010Williams}. 
Observations of 1ES 1028+511 at $z=0.361$ by 
\citet[][]{2006Steenbrugge} yield no convincing evidence for X-ray WHIM absorption.

What is perceived to be more disconcerting is the lack of detection of O~VII absorbers
at the redshifts of detected O~VI absorbers along some random lines of sight. 
This is because, overall, the O~VII line is predicted to be 
the most abundant
and anecdotal evidence suggests substantial coincidence between O~VI and O~VII \citep[e.g.,][]{2006bCen}. 
A statistically significant upper limit placed on the mean column density of O~VII absorbers 
at the locations of a sizeable set of detected O~VI absorbers using stacking techniques by \citet[][]{2009Yao} 
prompts them to call into question the very existence of the high temperature ($T\ge 10^6$K) portion of the
WHIM, although the limited sensitivity and
spectral resolution of the current X-ray observations may render any such conclusions less than definite.

Therefore, at this juncture, it is pressing
to statistically address this lack of significant coincidence between O~VI and O~VII absorbers
and other issues theoretically, through higher resolution simulations that are necessary in order to well
resolve the interfaces of multi-phase media.
This is the primary purpose of this paper.
We use two simulations of high resolution of $0.46h^{-1}$kpc 
and box size of $20-30h^{-1}$Mpc to perform much more detailed 
characterization of O~VI and O~VII lines to properly compare to extant observations.
This high resolution is to be compared with
 $83h^{-1}$kpc resolution in our previous simulations \citep[][]{2006Cen, 2006bCen},
$25-49h^{-1}$kpc resolution in \citet[][]{2011Smith} and \citet[][]{2011Shull}, 
$1.25-2.5h^{-1}$kpc in \citet[][]{2012Oppenheimer} 
and 
$1.25-2.5h^{-1}$kpc in \citet[][]{2011TepperGarcia}, 
resolves the Jeans scale of WHIM by 2-3 orders of magnitude and interfaces between 
gas phases of different temperatures in a multi-phase medium.
It is useful to distinguish, in the case of SPH simulations,
between the gravity force resolution and the resolution of the hydrodynamics solver,
with the latter being worse than the former by a factor of order a few.
It is also useful to keep in mind the initial cell size or interparticle separation, because in both SPH 
and adaptive mesh refinement (AMR) simulations not all regions are resolved by the maximum resolution. 
Calling this ``mean region resolution" $\Delta_{\rm root}$, 
$\Delta_{\rm root}=(117, 25-49, 125, 195)h^{-1}$kpc for 
[this paper, \citet[][]{2011Smith}, \citet[][]{2012Oppenheimer}, \citet[][]{2011TepperGarcia}].
We note that a region of overdensity $\delta$ is approximately resolved at a resolution of 
${\rm C}\Delta_{\rm root}\delta^{-1/3}$ (up to a pre-specified highest resolution),
where the pre-factor $C$ is about unity for AMR simulations and $\sim 2$ for SPH simulations.
Using Lagrangian SPH or AMR approaches becomes necessary for regions
$\delta\ge 300$, because simulations of a similar resolution with the uni-grid method become increasingly impractical
(largely due to limitations of computer memory).
A more important advantage with very high resolution simulations has to do with 
the need to resolve galaxies, which in turn allows for a more self-consistent treatment
of the feedback processes from star formation, namely, the temporal and spatial distribution of metals and energy 
deposition rates to the CGM and IGM and their effects on subsequent star formation.

Our new simulations, in agreement with earlier findings,
reaffirm quantitatively the existence of WHIM and furthermore show that the properties of 
the WHIM with respect to O~VI line and O~VI-O~VII relations are fully consistent with observations.
In particular, the observed upper limit of the mean coincidental O~VII column density of detected O~VI absorbers
is higher than what is predicted by the simulations by a factor of $\sim 2.5-4$.
Higher sensitivity X-ray observations or a larger sample by a factor of $\sim 10$ should 
test this prediction definitively.
The outline of this paper is as follows.
In \S 2.1 we detail simulation parameters and hydrodynamics code,
followed by a description of our method of making synthetic O~VI and O~VII spectra in \S 2.2, 
which is followed by a description of how we average the two separate simulations C (cluster) and V (void) run in \S 2.3.
Results are presented in \S 3.
In \S 3.1 we present some observables for O~VI to compare to observations 
to provide additional validation of the simulations.
In \S 3.2 we dissect the simulations to provide a physical 
analysis of the O~VI and O~VII absorbers.
In \S 3.3 results on the coincidences between O~VI and O~VII lines are given.
Conclusions are summarized in \S 4.

\section{Simulations}\label{sec: sims}

\subsection{Hydrocode and Simulation Parameters}

We perform cosmological simulations with the AMR Eulerian hydro code, Enzo 
\citep[][]{1999aBryan, 1999bBryan, 2005OShea, 2009Joung}.  
First we ran a low resolution simulation with a periodic box of $120~h^{-1}$Mpc on a side.
We identified two regions separately, one centered on
a cluster of mass of $\sim 2\times 10^{14}\msun$
and the other centered on a void region at $z=0$.
We then resimulate each of the two regions separately with high resolution, but embedded
in the outer $120h^{-1}$Mpc box to properly take into account large-scale tidal field
and appropriate boundary conditions at the surface of the refined region.
We name the simulation centered on the cluster ``C" run
and the one centered on the void  ``V" run.
The refined region for ``C" run has a size of $21\times 24\times 20h^{-3}$Mpc$^3$
and that for ``V" run is $31\times 31\times 35h^{-3}$Mpc$^3$.
At their respective volumes, they represent $1.8\sigma$ and $-1.0\sigma$ fluctuations.
{\color{burntorange}\bf
The root grid has a size of $128^3$ with $128^3$ dark matter particles.
The initial static grids in the two refined boxes
correspond to a $1024^3$ grid on the outer box.
The initial number of dark matter particles in the two refined boxes
correspond to $1024^3$ particles on the outer box.
This translates to initial condition in the refined region having a mean interparticle-separation of 
$117h^{-1}$kpc comoving and dark matter particle mass of $1.07\times 10^8h^{-1}\msun$.
}
The refined region is surrounded by two layers (each of $\sim 1h^{-1}$Mpc) 
of buffer zones with 
particle masses successively larger by a factor of $8$ for each layer, 
which then connects with
the outer root grid that has a dark matter particle mass $8^3$ times that in the refined region.
The initial density fluctuations 
are included up to the Nyquist frequency in the refined region.
The surrounding volume outside the refined region 
is aso followed hydrodynamically, which is important in order to properly capture
matter and energy exchanges at the boundaries of the refined region.
Because we still can not run a very large volume simulation with adequate resolution and physics,
we choose these two runs of moderate volumes to represent two opposite environments that possibly bracket the average.

We choose the mesh refinement criterion such that the resolution is 
always better than $460h^{-1}$pc physical, corresponding to a maximum
mesh refinement level of $11$ at $z=0$.
The simulations include
a metagalactic UV background
\citep[][]{2012Haardt},  
and a model for shielding of UV radiation by neutral hydrogen 
\citep[][]{2005Cen}.
The simulations also include metallicity-dependent radiative cooling and heating \citep[][]{1995Cen}. 
We clarify that our group has included metal cooling and metal heating (due to photoionization of metals) 
in all our studies since \citet[][]{1995Cen},
contrary to some claims \citep[e.g.,][]{2009Wiersma, 2011TepperGarcia}.
Star particles are created in cells that satisfy a set of criteria for 
star formation proposed by \citet[][]{1992CenOstriker}.
Each star particle is tagged with its initial mass, creation time, and metallicity; 
star particles typically have masses of $\sim$$10^6\msun$.

Supernova feedback from star formation is modeled following \citet[][]{2005Cen}.
Feedback energy and ejected metal-enriched mass are distributed into 
27 local gas cells centered at the star particle in question, 
weighted by the specific volume of each cell (i.e., weighting is equal to the inverse of density), 
which is to mimic the physical process of supernova
blastwave propagation that tends to channel energy, momentum and mass into the least dense regions
(with the least resistance and cooling).
We allow the whole feedback processes to be hydrodynamically coupled to surroundings
and subject to relevant physical processes, such as cooling and heating, as in nature.
The extremely inhomogeneous metal enrichment process
demands that both metals and energy (and momentum) are correctly modeled so that they
are transported into right directions in a physically sound (albeit still approximate 
at the current resolution) way, at least in a statistical sense.

\begin{figure}[h!]
\centering
\hskip -1.0cm
\resizebox{3.5in}{!}{\includegraphics[angle=0]{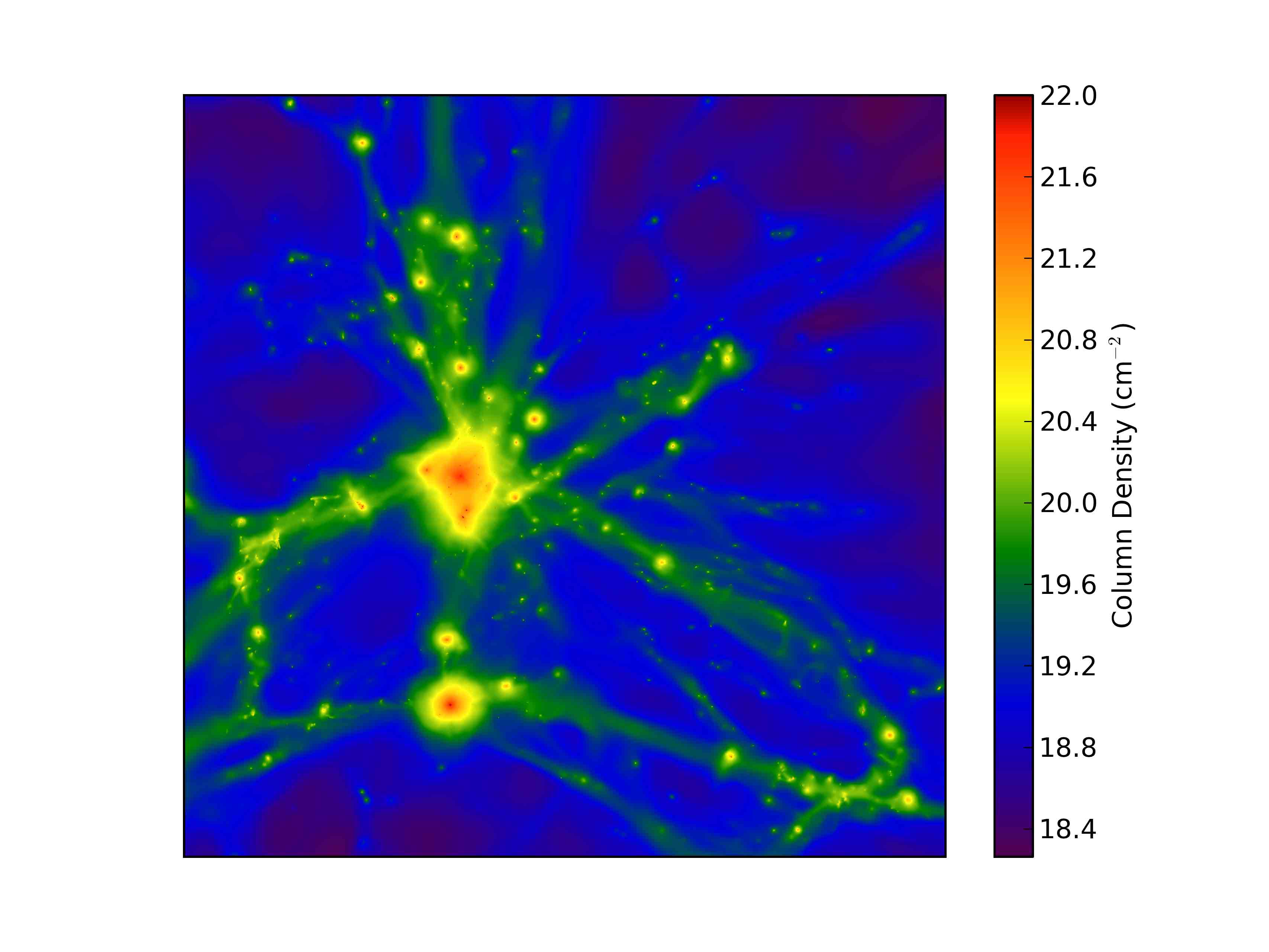}}
\hskip -1.0cm
\resizebox{3.5in}{!}{\includegraphics[angle=0]{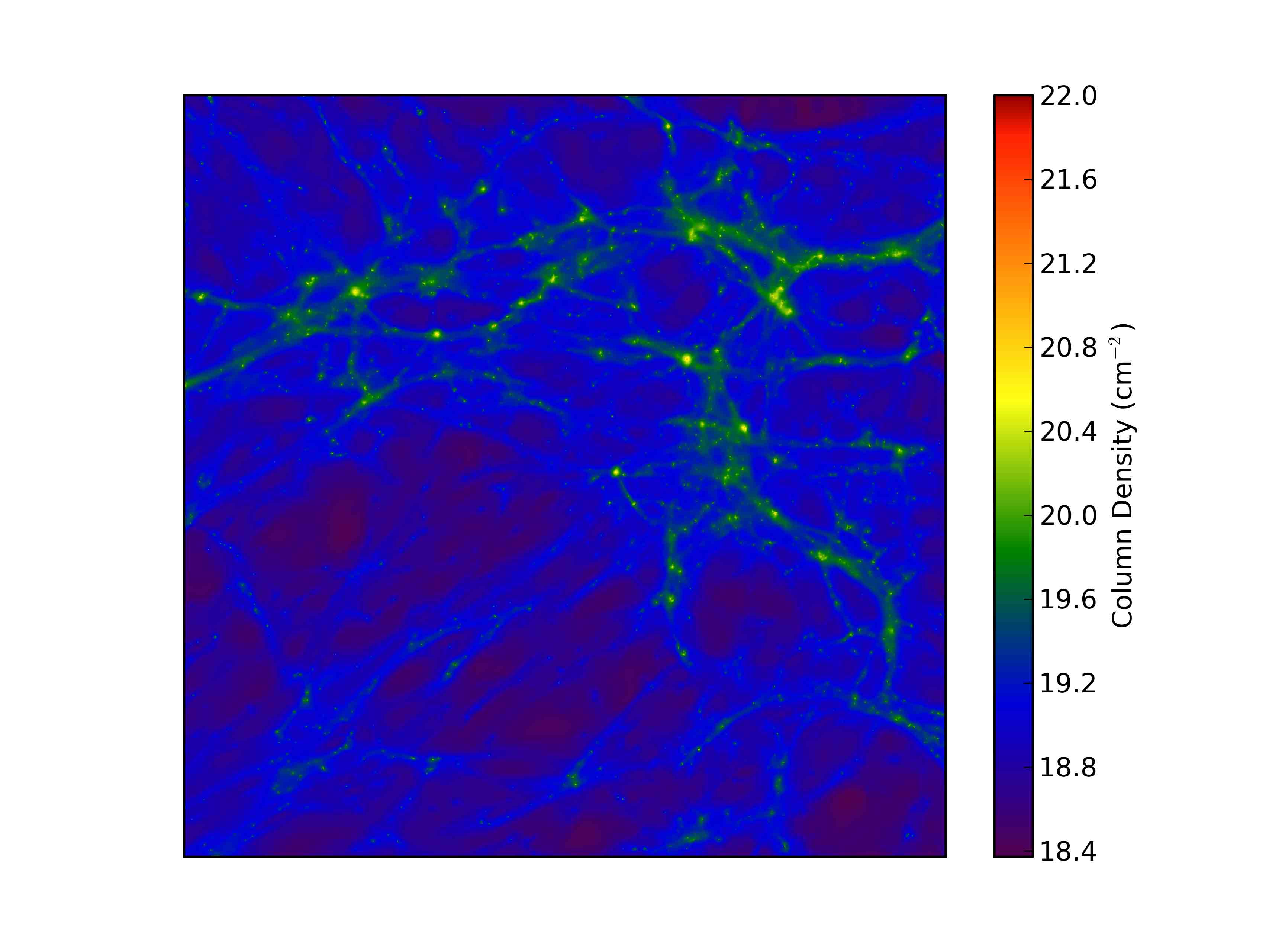}}
\vskip 0.01cm
\hskip -1.0cm
\resizebox{3.5in}{!}{\includegraphics[angle=0]{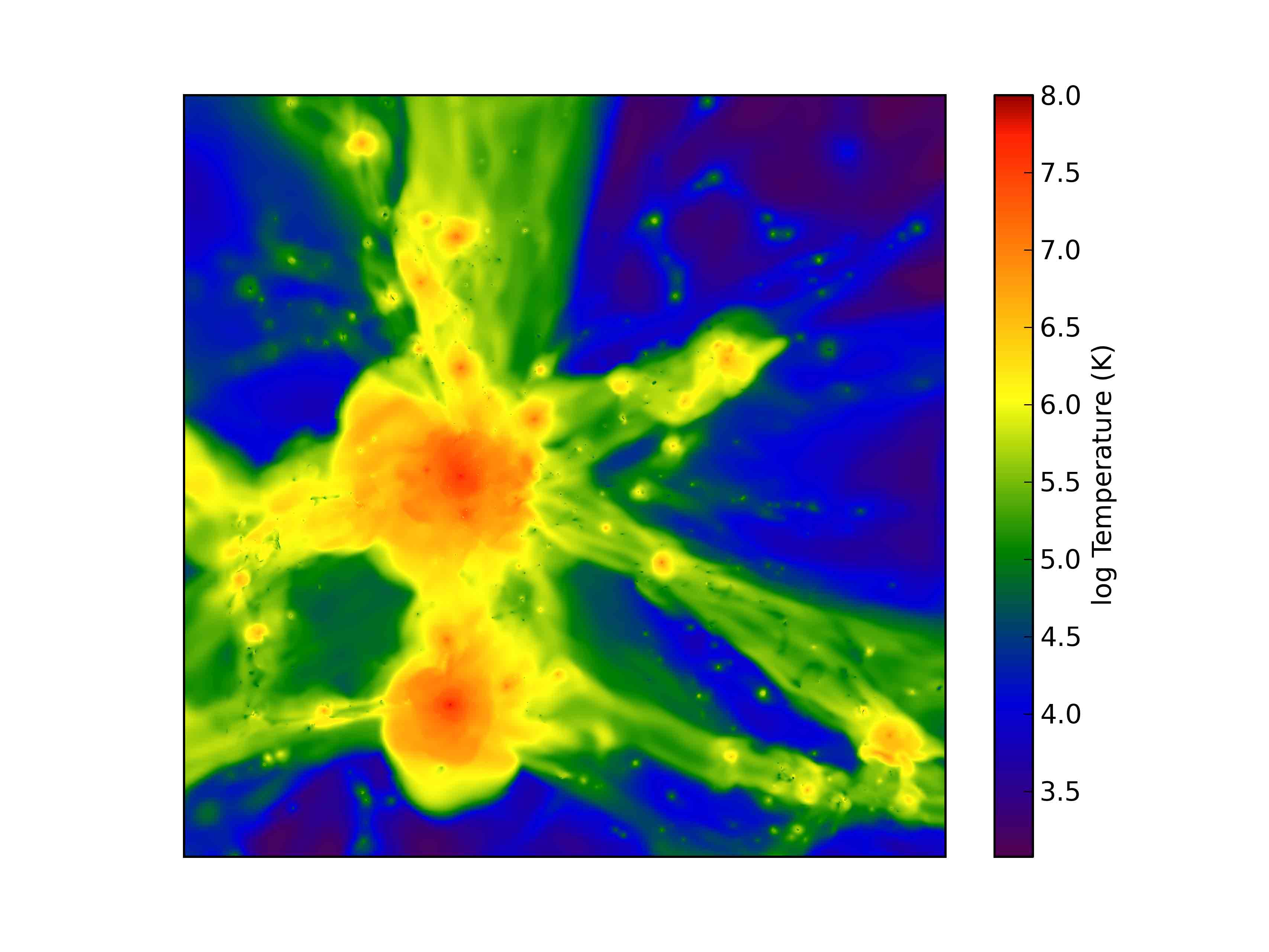}}
\hskip -1.0cm
\resizebox{3.5in}{!}{\includegraphics[angle=0]{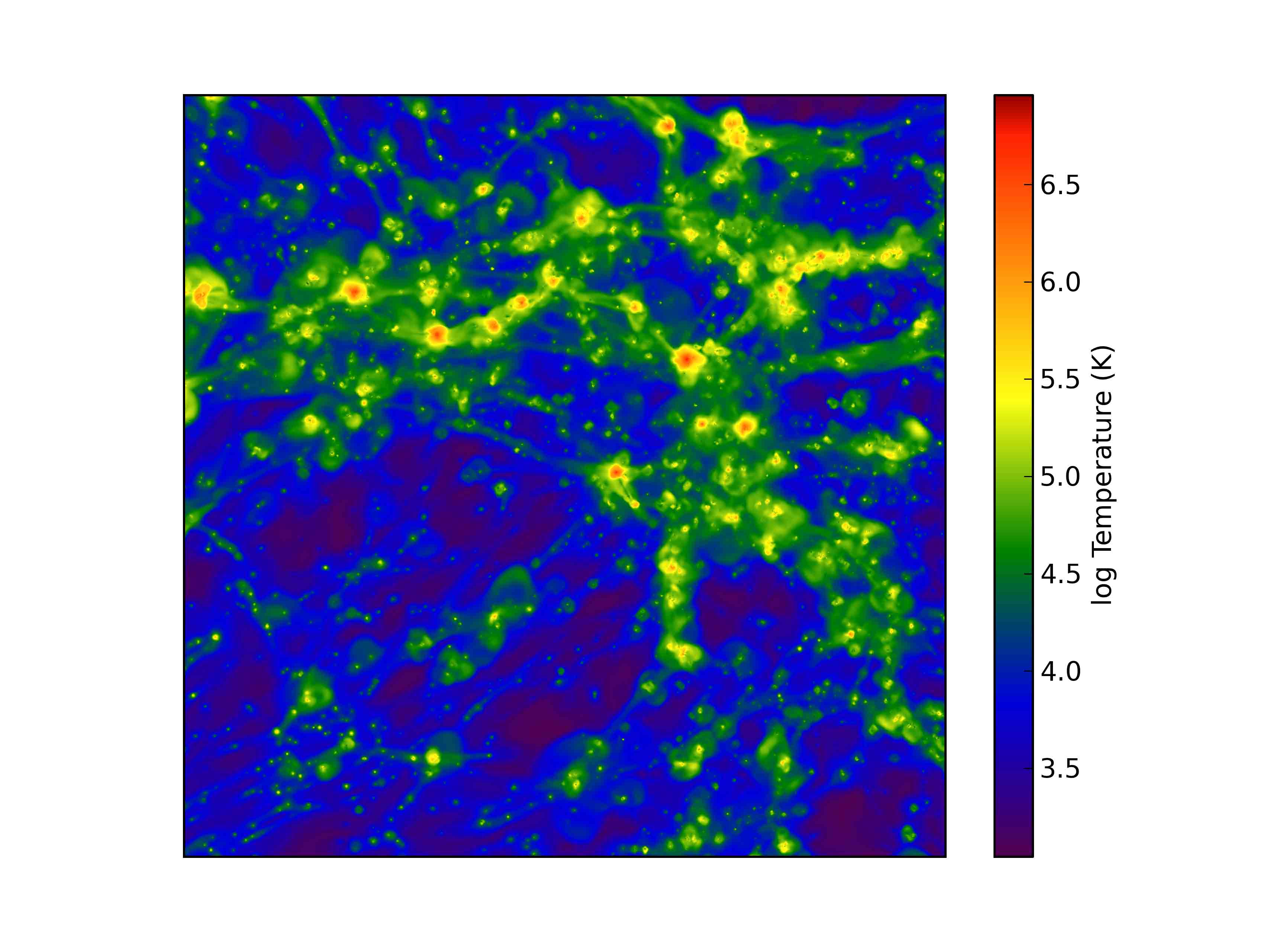}}
\caption{
Top-left and bottom-left panels show the gas density and density-weighted temperature 
projection of a portion of the refinement box of the C run of size ($18h^{-1}$Mpc)$^3$.
Top-right and bottom-right panels show the gas density and density-weighted temperature 
projection of a portion of the refinement box of the V run of size ($30h^{-1}$Mpc)$^3$.
}
\label{fig:sim}
\end{figure}

The primary advantages of this supernova energy based feedback mechanism are three-fold.
First, nature does drive winds in this way and energy input is realistic.
Second, it has only one free parameter $e_{SN}$, namely, the fraction of the rest mass energy of stars formed
that is deposited as thermal energy on the cell scale at the location of supernovae.
Third, the processes are treated physically, obeying their respective conservation laws (where they apply),
allowing transport of metals, mass, energy and momentum to be treated self-consistently 
and taking into account relevant heating/cooling processes at all times.
We use $e_{SN}=1\times 10^{-5}$ in these simulations.
The total amount of explosion kinetic energy from Type II supernovae
with a Chabrier IMF translates to $e_{SN}=6.6\times 10^{-6}$.
Observations of local starburst galaxies indicate
that nearly all of the star formation produced kinetic energy (due to Type II supernovae)
is used to power galactic superwinds \citep[e.g.,][]{2001Heckman}.
Given the uncertainties on the evolution of IMF with redshift (i.e., possibly more top heavy at higher redshift)
and the fact that newly discovered prompt Type I supernovae contribute a comparable
amount of energy compared to Type II supernovae, it seems that our adopted value for
$e_{SN}$ is consistent with observations and within physical plausibility.
Test of the success of this feedback approach comes empirically.
As we show in \citet[][]{2012Cen}, the metal distribution in and around galaxies over a wide range of redshift
is in good agreement with respect to the properties of observed damped $\lya$ systems;
this is a non-trivial success and provides strong validation of the simulations.
We will provide additional validation of the simulations in \S 3.1.

To better understand differences in results between AMR and SPH simulations that we will discuss later,
we note here that the evolution of metals in the two types of simulations is treated rather differently.
In AMR simulations metals are followed hydrodynamically by solving 
the metal density continuity equation with sources (from star formation feedback) and sinks (due to subsequent star formation),
whereas in SPH simulations of WHIM one does not separately solve the metal density continuity equation.
Thus, metal mixing and diffusion through advection, turbulence and other hydrodynamic processes
are properly captured in AMR simulations.
While some SPH simulations have implemented metal diffusion schemes that are motivated by 
some subgrid turbulence model as a remedy parameterized 
to roughly match results from hydrodynamic simulations \citep[e.g.,][]{2010Shen},
most SPH simulations of WHIM obtain gas metallicities based on kernel-smoothed 
metal masses of feedback SPH particles that are assigned at birth and un-evolved
\citep[e.g.,][]{2011TepperGarcia, 2009Oppenheimer, 2012Oppenheimer}.
In the simulations of \citet[][]{2012Oppenheimer} ``feedback" SPH particles with initially given metal masses 
are launched (in random directions) to be transported ballistically to sufficiently large distance ($\sim 10$kpc),
after allowance for some period of hydrodynamic de-coupling between the feedback SPH particles and other neighboring SPH particles.
Once some of the feedback parameters are fixed, this approach produces definitive predictions 
with respect to various aspects of stellar and IGM metallicity and others
\citep[e.g.,][]{2003Springel, 2009Oppenheimer, 2010Tornatore, 2011Dave, 2011bDave, 2012Oppenheimer}. 
It is likely that mixing of metals on small to intermediate scales ($\sim 1-100$kpc) in SPH simulations
\citep[e.g.,][]{2012Oppenheimer, 2011TepperGarcia} is probably substantially underestimated.
This significant difference in the treatment of metal evolution may have contributed, in a large part,
to some discrepancies between SPH and AMR hydrodynamic simulations, as we will discuss later.

We use the following cosmological parameters that are consistent with 
the WMAP7-normalized \citep[][]{2010Komatsu} LCDM model:
$\Omega_M=0.28$, $\Omega_b=0.046$, $\Omega_{\Lambda}=0.72$, $\sigma_8=0.82$,
$H_0=100 h {\rm km s}^{-1} {\rm Mpc}^{-1} = 70 {\rm km} s^{-1} {\rm Mpc}^{-1}$ and $n=0.96$.

Figure~\ref{fig:sim} shows the density and temperature fields of the two simulations.
The environmental contrast between the two simulations is evident.
We also note that there is substantial overlap visually between the two simulations 
in that both cover the ``field" environment, which we have shown quantitatively in \citet[][]{2011Tonnesen}.
In other words, these two simulations cover two extreme environments - voids and clusters -
with substantial overlap of intermediate environment that facilitates 
possible averaging of some computed quantities, with proper normalizations by independent observational constraints.

\subsection{Generation of Synthetic O~VI and O~VII Absorption Lines}

The photoionization code CLOUDY (Ferland et al. 1998) is used 
post-simulation to compute the abundance of O~VI and O~VII, adopting the shape of the UV
background calculated by \citet[][]{2012Haardt} normalized by the
intensity at 1 Ryd determined by 
\citet[][]{1999Shull} and assuming ionization equilibrium.

\begin{figure}[h!]
\vskip -0.0cm
\centering
\hskip -0.2in
\resizebox{3.41in}{!}{\includegraphics[angle=0,height=3in, width=2.8in]{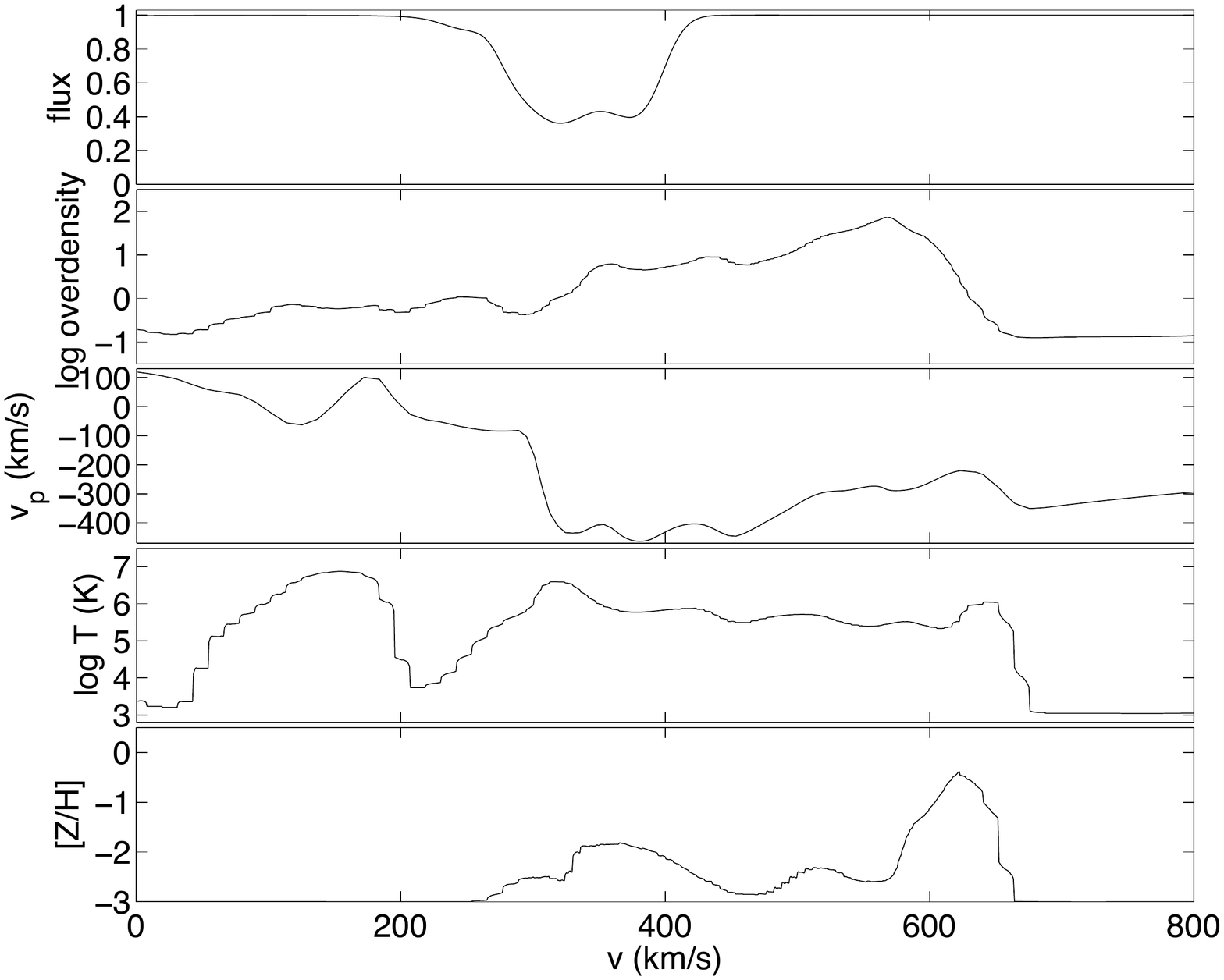}}
\hskip -0.2in
\resizebox{3.41in}{!}{\includegraphics[angle=0,height=3in, width=2.8in]{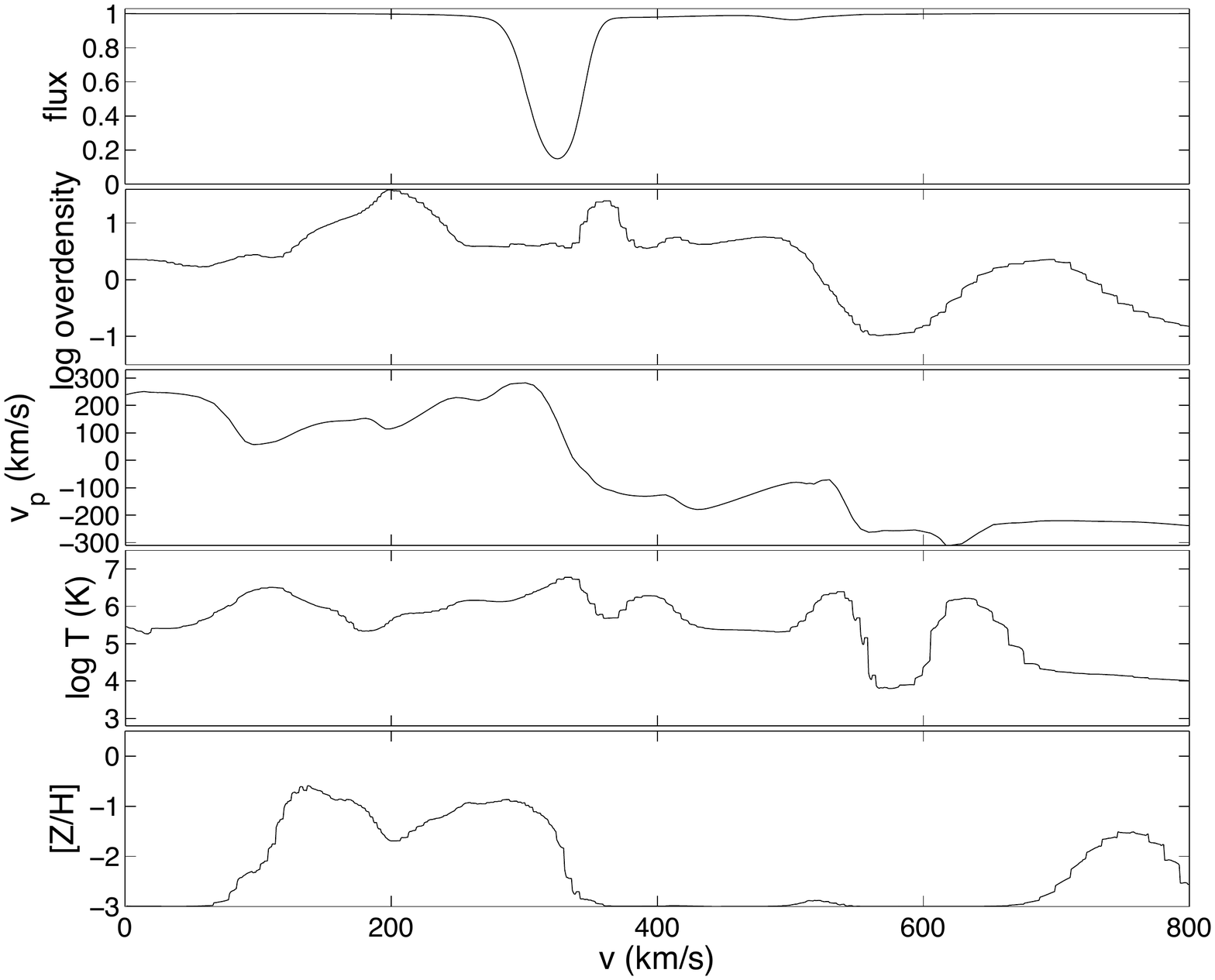}}
\vskip -0.5cm
\caption{
shows flux spectra of two separate O~VI lines and physical conditions.
The left and right cases have column densities of $\log {\rm N(OVI)cm^{2}}=14.48$ and $14.30$, respectively.
The five panels from top to bottom are:
flux, gas overdensity, proper peculiar velocity, temperature and metallicity in solar units.
While the x-axis for the top panel is the Hubble velocity, the x-axis for the bottom four panels
is physical distance that is multiplied by the Hubble constant.
}
\label{fig:spectra}
\end{figure}

We generate synthetic absorption spectra using a code similar to that used
in our earlier papers \citep[e.g.,][]{1994Cen, 2001Cen, 2006bCen},
given the density, temperature, metallicity and velocity fields from simulations.
Each absorption line is identified by the interval between one downward and the next upward crossing 
in the synthetic flux spectrum without noise at a flux equal to $0.99$ (flux equal to unity corresponds to an unabsorbed continuum).
Since the absorption lines in question are sparsely distributed in velocity space,
their identifications have no significant ambiguity.
Column density, equivalent width, Doppler width,
mean column density weighted velocity and physical space locations, 
mean column density weighted temperature, density and metallicity are computed for each line.
We sample the C and V run, respectively, with $72,000$ and $168,000$ 
random lines of sight at $z=0$, 
with a total pathlength of $\Delta z\sim 2000$.
While a detailed Voigt profile fitting of the flux spectrum would have enabled 
closer comparisons with observations, 
simulations suggest that such an exercise does not necessarily provide a more clarifying 
physical understanding of the absorber properties, because 
bulk velocities are very important (see Figure~\ref{fig:bT_O6} below) 
and velocity substructures within an absorber
do not necessarily correspond to separate physical entities.

A small number of simulated spectra
may not serve to illustrate the extreme rich and complex physics involved.
It may even be misleading in the sense that any statistical conclusions
drawn based on anecdotal evidence could be substantially wrong.
Thus, we will present two absorption spectrum segments merely only for the purpose of illustration.
Figure~\ref{fig:spectra} shows two O~VI lines and their associated physical environment.

\begin{figure}[ht]
\vskip -0.5cm
\centering
\resizebox{5.0in}{!}{\includegraphics[angle=0]{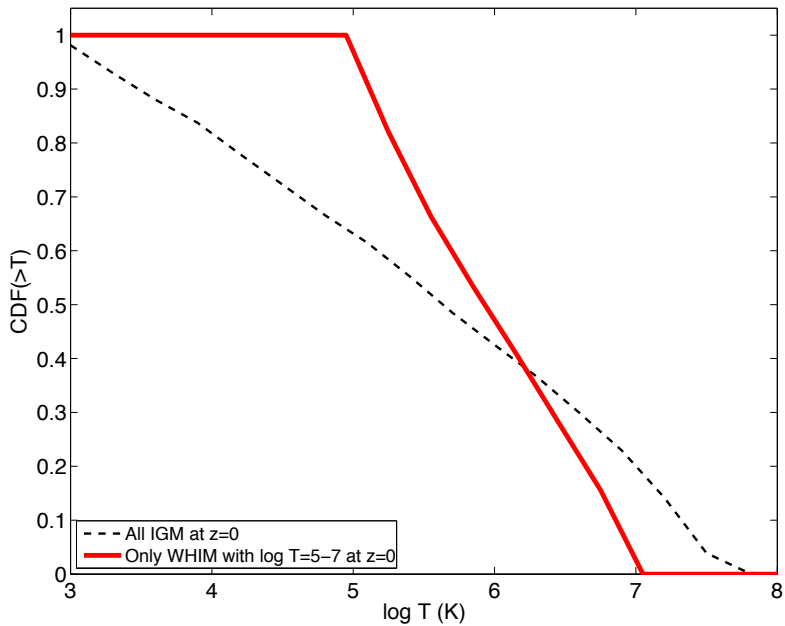}}
\vskip -0.4cm
\caption{
shows the cumulative probability distribution function (CDF) of the IGM at $z=0$ as a function of gas temperature 
(black dashed curve) and that of the WHIM only in the temperature range $T=10^5-10^7$K (red solid curve); 
stars are not included.
}
\label{fig:This}
\end{figure}

\subsection{Averaging C and V Runs}

The C and V runs at $z=0$ are used to obtain an ``average" of the universe.
This cannot be done precisely without much larger simulation volumes, which is presently not feasible.
Nevertheless, it is still possible to obtain an approximate average.
Since the WHIM is mostly closely associated with groups and clusters of galaxies,
we will use X-ray clusters as an appropriate ``normalization" anchor point.
We normalize averaging weightings of the C and V runs by requiring that
the fraction of hot gas with temperature $T\ge 10^7$K is consistent with 
the observed value of $\sim 15\%$ of baryons at $z=0$ \citep[][]{2011Bahcall}.
Note that small variations on the adopted X-ray gas fraction do not cause large changes in most of the results.
For comparative measures such as the coincidence rates between O~VI and O~VII absorbers,
the dependence on the normalization procedure is still weaker.
The results are shown in Figure~\ref{fig:This}, which shows the temperature distribution of entire IGM and WHIM at $z=0$.
In agreement with previous simulations \citep[e.g.,][]{1999Cen, 2001Dave, 2006Cen},
we find that $\sim 40\%$ of the IGM at $z=0$ is in WHIM.
This is compared to 35-40\% in \citet[][]{2011Smith}, 
24\% in \citet[][]{2010Dave} (limited to overdensities outside halos),
40\% in \citet[][]{2010Shen} 
and 
40\% and 50\% in \citet[][]{2010Tornatore} in their wind and black hole feedback models, respectievely.
In simulations of \citet[][]{1995Cen, 1999Cen, 2001Cen, 2006Cen, 2011Cen},
\citet[][]{2009Wiersma, 2011TepperGarcia}, \citet[][]{2010Shen} and \citet[][]{2011Shull},
in additional to radiative processes of a primordial gas,
both metal cooling (due to collisional excitation and recombination)
and metal heating (due to photo-ionization heating of metal species) in the presence of UV-X-ray background
are included, 
whereas in \citet[][]{2009Oppenheimer}, \citet[][]{2012Oppenheimer} and \citet[][]{2010Tornatore} 
only metal cooling is included.
\citet[][]{2011TepperGarcia} suggest that 
the relatively overall low fraction of WHIM in the latter (25\%) versus higher fraction 
in the former (35-50\%) may be accounted by the difference in the treatment of metal heating;
we concur with their explanation for at least part of the difference.
All these simulations have a box size of $\sim 50h^{-1}$Mpc, which still suffers from 
significant cosmic variance: 
\citet[][]{2001Dave} show that WHIM fraction increases from 37\% to 42\% from
a box size of $50h^{-1}$Mpc to $100h^{-1}$Mpc in two Eulerian simulations.
The amplitude of power spectrum has a similar effect and may be able to, at least in part,
account for some of the differences among the simulations;
$\sigma_8=(0.82,0.82,0.74,0.77,0.80,0.82)$ in 
[this work, \citet[][]{2011Shull}, \citet[][]{2011TepperGarcia}, \citet[][]{2010Shen}, \citet[][]{2010Tornatore}, \citet[][]{2012Oppenheimer}].
Gravitational collapse of longer waves powers heating of the IGM at later times.
We suggest that the peak of WHIM fraction at $z\sim 0.5$ found   
in the $25h^{-1}$Mpc simulation boxes in \citet[][]{2011Smith} is 
because of the small box size; in other words, available, reduced gravitational heat input
in the absence of breaking density waves of lengths longer than $25h^{-1}$Mpc 
at $z\le 0.5$ fails to balance the cooling due to (primarily) universal expansion
and (in part) radiative cooling.
This explanation is supported by the behavior of their simulation boxes of size
$50^{-1}$Mpc at low redshift ($z\le 0.5$).
Figure~\ref{fig:This} shows that 
within the WHIM temperature range, roughly equal amounts 
are at $T=10^5-10^6$K and $T=10^6-10^7$K.

\section{Results}

\subsection{Simulation Validation with Properties of O~VI Absorbers}

\begin{figure}[H]  
\vskip -0.7cm
\centering
\resizebox{5.0in}{!}{\includegraphics[angle=0]{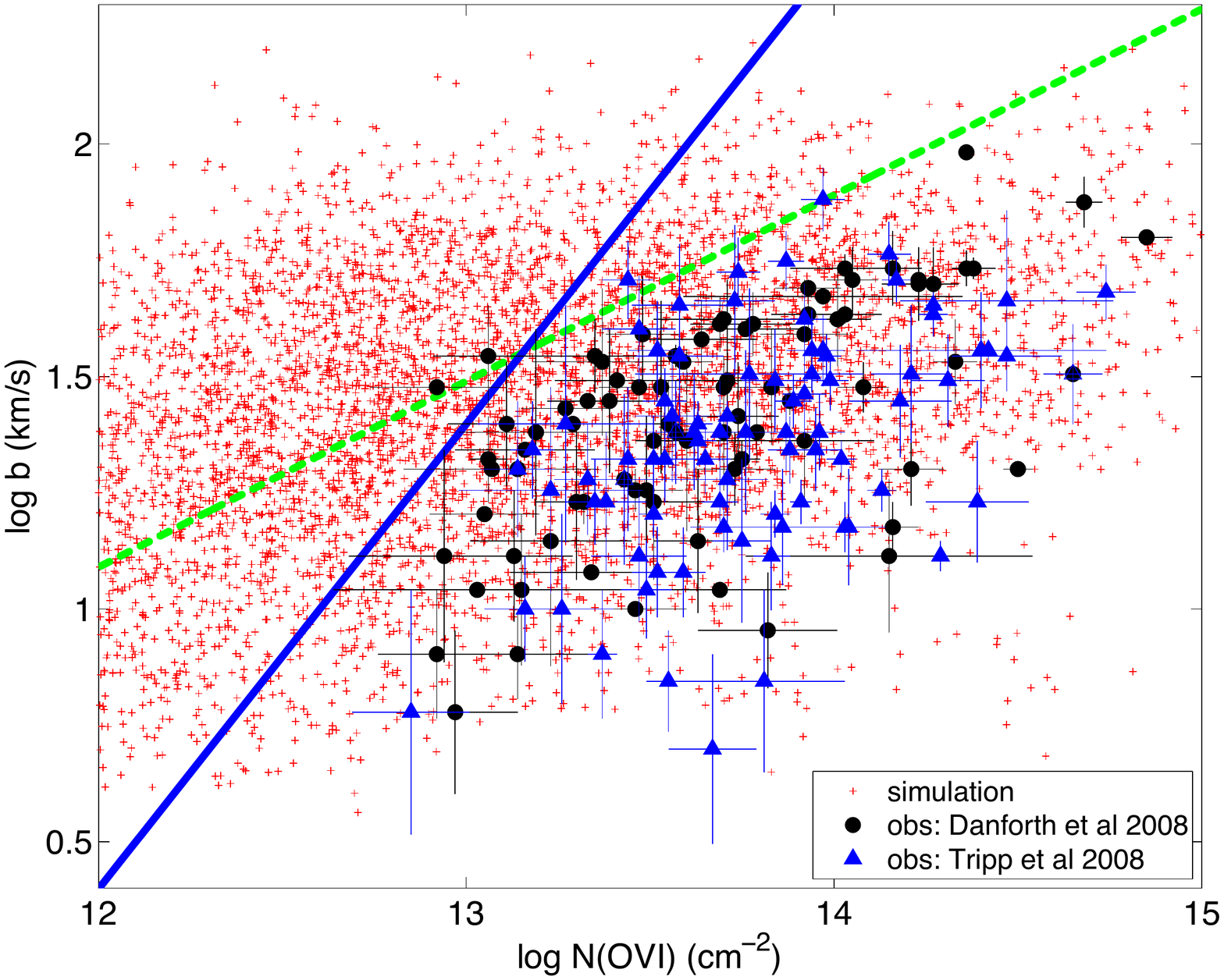}}
\vskip -1.0cm
\resizebox{5.0in}{!}{\includegraphics[angle=0]{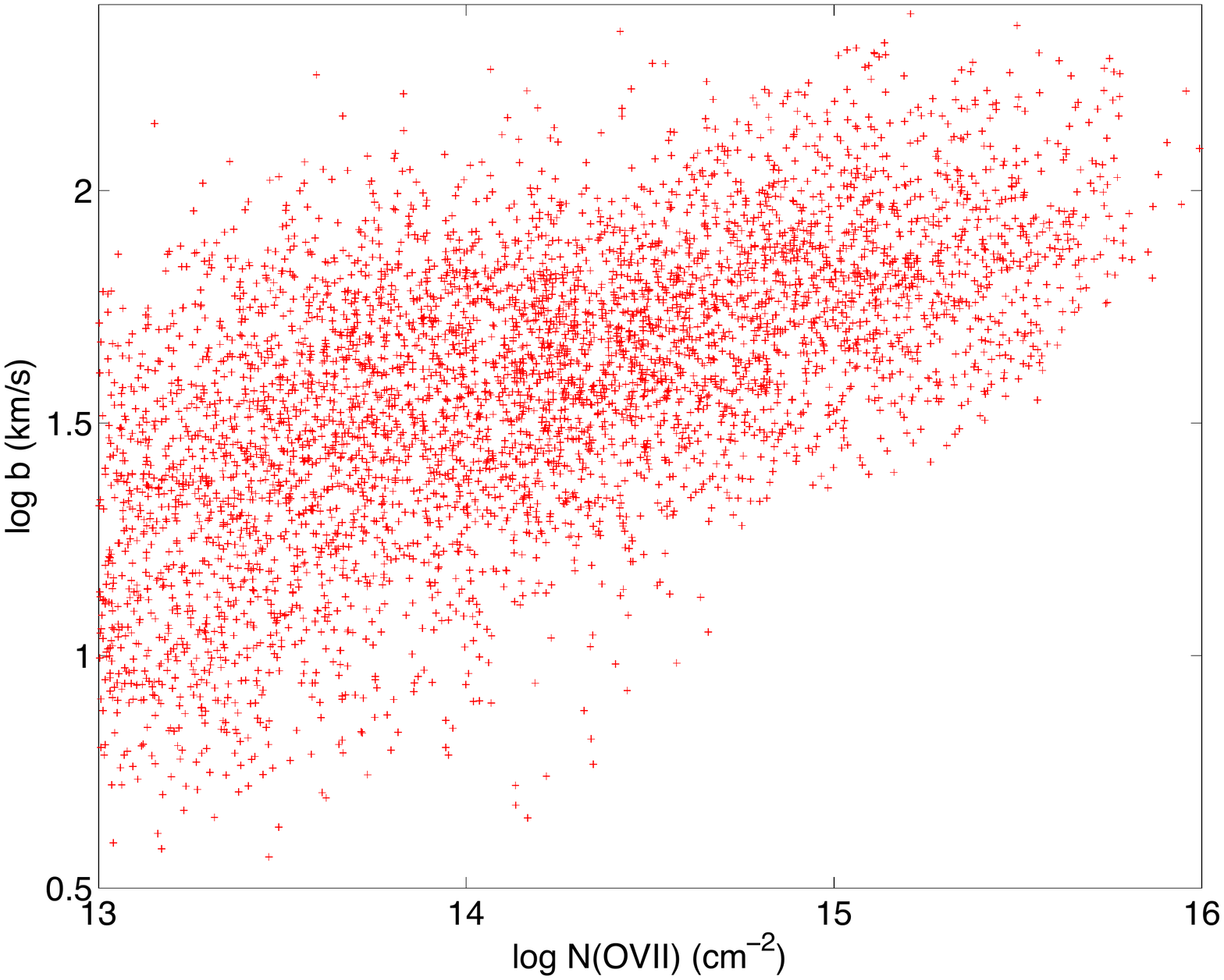}}
\vskip -0.4cm
\caption{
Top panel shows a scatter plot of simulated O~VI absorbers (red pluses) in the $b$-${\rm N(OVI)}$ plane.
Also shown as black dots and blue triangles are the observations from
 \citet[][]{2008Danforth} and \citet[][]{2008Tripp}, respectively.
The green dashed line of slope $4/10$ is only intended to guide the eye to suggest that there appears to be
a desert of observed O~VI absorbers in the upper left corner.
The blue solid line of unity logarithmic slope is a requirement 
for the decrement at the flux trough of the weaker of the O~VI doublet to be 4\%:
${\rm b=25(N(OVI)/10^{13}cm^{-2})}\kms$.
Bottom panel shows the same for the O~VII absorbers.
}
\label{fig:bN_O6}
\end{figure}

The present simulations have been shown to produce 
the metal distribution in and around galaxies over a wide range of redshift ($z=0-4$)
that is in good agreement with 
respect to the properties of observed damped $\lya$ systems
\citep[][]{2012Cen}. 
Here we provide additional, more pertinent validation with respect to O~VI absorbers in the IGM at $z=0$.
The top panel of Figure~\ref{fig:bN_O6} shows a scatter plot 
of simulated O~VI absorbers (red pluses) in the Doppler width ($b$)-O~VI column density 
[${\rm N(OVI)}$] plane, compared to observations.
The agreement is excellent in that the observed O~VI absorbers occupy a region that
overlaps with the simulated one.
It is intriguing to note that the simulations predict a large number
of large $b$, low ${\rm N(OVI)}$ (i.e., broad and shallow) absorbers
in the region ${\rm b>31 (N(OVI)}/10^{14}{\rm cm^{-2}})^{0.4}$km/s, 
corresponding to the upper left corner to the green dashed line,
where there is no observed O~VI absorber.
This green dashed line, however, has no physical meaning to the best of our knowledge.
The blue solid line of unity logarithmic slope has a clear physical origin, which is a requirement 
for the decrement at the flux trough of the weaker of the O~VI doublet to be 4\%: ${\rm b=25(N(OVI)/10^{13}cm^{-2})}\kms$.
Current observational data are heterogeneous with varying qualities.
Thus, the blue solid line is a much simplified characterization of the complex situation.
Nevertheless, one could understand the desert of observed O~VI absorbers in the upper left corner to the blue solid line,
thanks to the difficulty of identifying broad and shallow lines in existing observations.
We attribute the ``missing" observed O~VI lines in the upper right corner
between the blue solid line and the green dashed line, in part, to the observational procedure
of Voigt profile fitting that may break up some large $b$ lines into separate, narrower components,
whereas no such procedure is performed in the presented simulation results.

Ongoing and upcoming observations by the Cosmic Origins Spectrograph (COS) 
\citep[e.g.,][]{2009Froning, 2009Shull, 2012Green} will be able to substantially improve in sensitivity and thus 
likely be able to detect a sizeable number of O~VI lines in the upper left corner to the blue solid line.
Quantitative distribution functions of $b$ parameter 
will be shown in Figure~\ref{fig:bhis} later, for which COS may provide a strong test.
The bottom panel of Figure~\ref{fig:bN_O6} shows a scatter plot of simulated O~VII absorbers (red pluses);
because there is no data to compare to, we only note that the positive correlation between $b$  
and ${\rm N(OVI)}$ is stronger for O~VII lines than for O~VI lines,
in part due to less important contribution to the O~VII lines from photoionization
and in part due to positive correlation between density and velocity dispersion.

\begin{figure}[h!]
\hskip -0.7cm
\centering
\resizebox{6.0in}{!}{\includegraphics[angle=0]{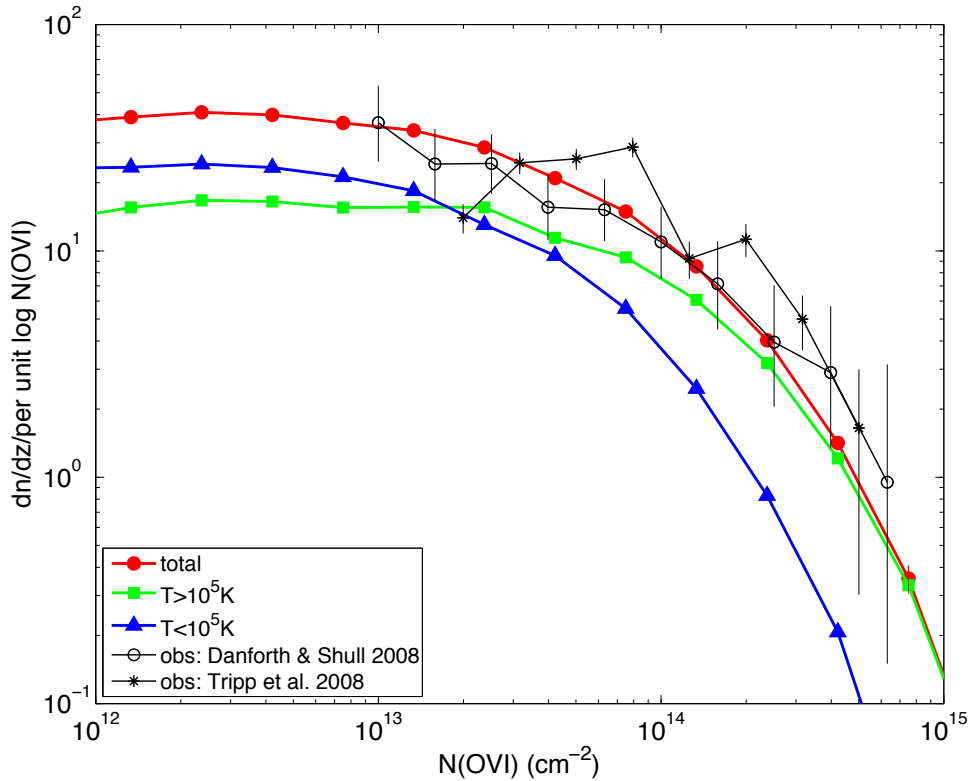}}
\vskip -0.4cm
\caption{
shows the O~VI line density as a function of column density,
defined to be the number of lines per unit redshift per unit logarithmic interval of the column density.
The red solid dots, green squares and blue triangles
are the total, collisionally ionized and photoionized absorbers, respectively.
Also shown as black open circles and stars are the observations from
 \citet[][]{2008Danforth} and \citet[][]{2008Tripp}, respectively.
}
\label{fig:dndz_O6}
\end{figure}

Figure~\ref{fig:dndz_O6} shows O~VI line density as a function of column density.
The agreement between simulations and observations of 
\citet[][]{2008Danforth} is excellent over the entire column density,
${\rm N(OVI)}\sim 10^{13}-10^{15}$cm$^{-2}$, where comparison can be made.
The simulation results are up to a factor of $\sim 2$ below the 
observational results of \citet[][]{2008Tripp} 
in the column density range ${\rm N(OVI)}\sim 10^{13.7}-10^{14.5}$cm$^{-2}$.
Some of the disagrement is due to different treatments in defining lines in that 
we do not perform Voigt profile fitting thus deblending of non-gaussian profiles
into multiple components, where the observational groups do 
and different groups often impose different, subjective criteria of 
choosing the ``goodness" of the fit.
The down turn of line density towards lower column densities from 
${\rm N(OVI)}\sim 10^{13.9}$cm$^{-2}$ from \citet[][]{2008Tripp} 
as well as the lower values in the column density range ${\rm N(OVI)}\sim 10^{13.2}-10^{13.7}$cm$^{-2}$
of \citet[][]{2008Danforth}
may be related to the ``missing" broad and shallow lines, as indicated 
in the top panel of Figure~\ref{fig:bN_O6}.
It is noted that the observed line density at
${\rm N(OVI)}\sim 10^{13}$cm$^{-2}$ of the \citet[][]{2008Danforth} data
displays an upturn and lies on top of the simulated curve.
Closer examination reveals that this is due to the presence of two relatively broad absorbers at 
${\rm N(OVI)}\sim 10^{13}$cm$^{-2}$ and $b\sim 30\kms$.
We expect that the upcoming COS observations will substantially raise
the line density at ${\rm N(OVI)}\le 10^{13.5}$cm$^{-2}$.

These results show that our simulation results are realistic with respect to the abundance of O~VI lines in the 
CGM and IGM.
This is a substantial validation of the simulations, when considered in conjunction with
the success of the simulations with respect to the damped $\lya$ systems \citep[][]{2012Cen}. 
The damped $\lya$ systems primarily originate in gas within the virial radii of galaxies, 
whereas the O~VI absorbers examined here extend well into the IGM,
some reaching as far as the mean density of the universe (see Figure~\ref{fig:Toverd_O6} below).
In combination, they require the simulations to have substantially correctly modeled
the propagation of initial metal-enriched blastwaves from sub-kpc scales to 
hundreds of kiloparsecs as well as other complex thermodynamics, at least in a statistical sense.
Since O~VII absorbers arise in regions in-between,
this gives us confidence that O~VII lines are also modeled correctly and the comparisons that we will make
between O~VI and O~VII lines are meaningful.

\begin{figure}[H]
\centering
\resizebox{5.0in}{!}{\includegraphics[angle=0]{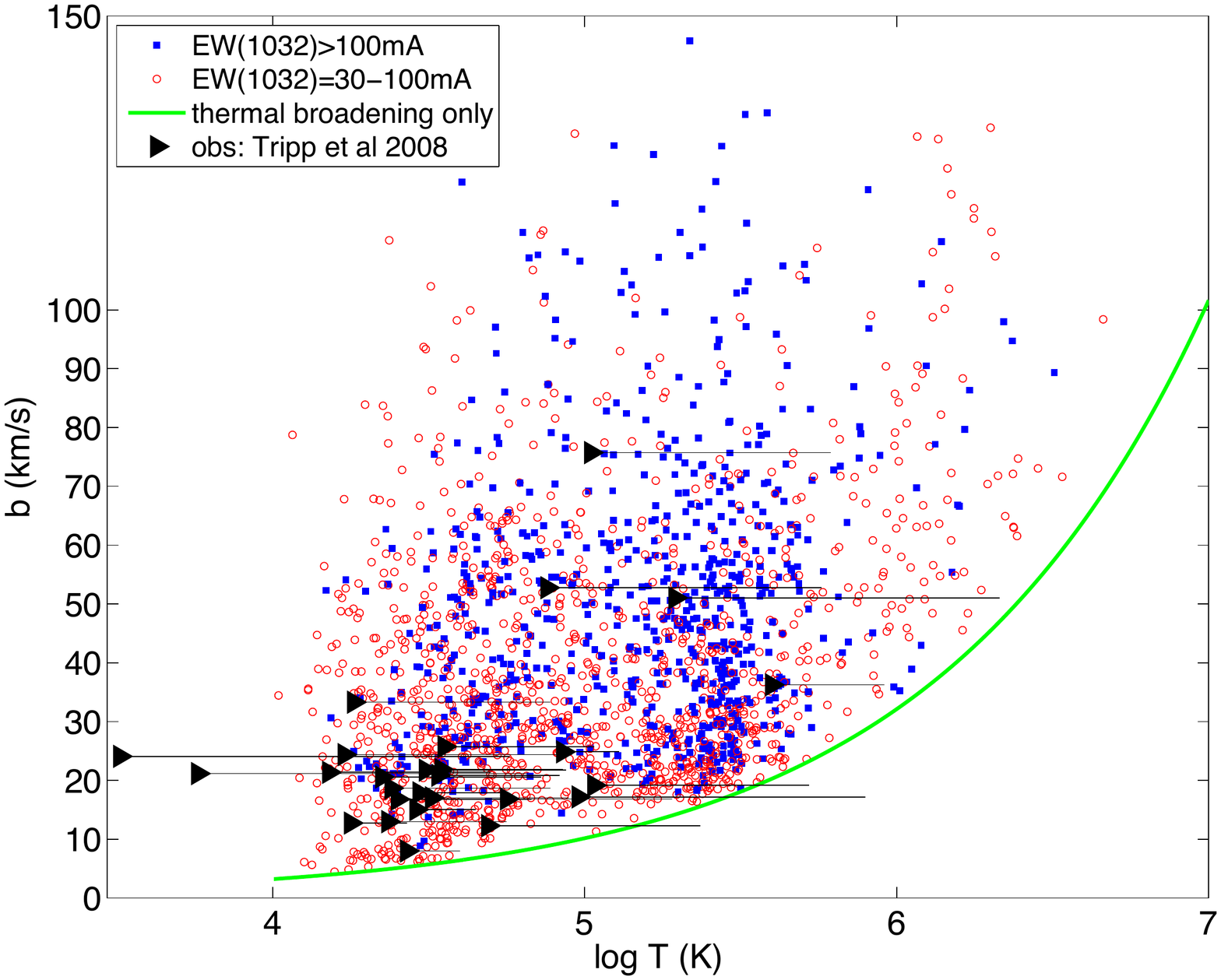}}
\vskip -1.0cm
\resizebox{5.0in}{!}{\includegraphics[angle=0]{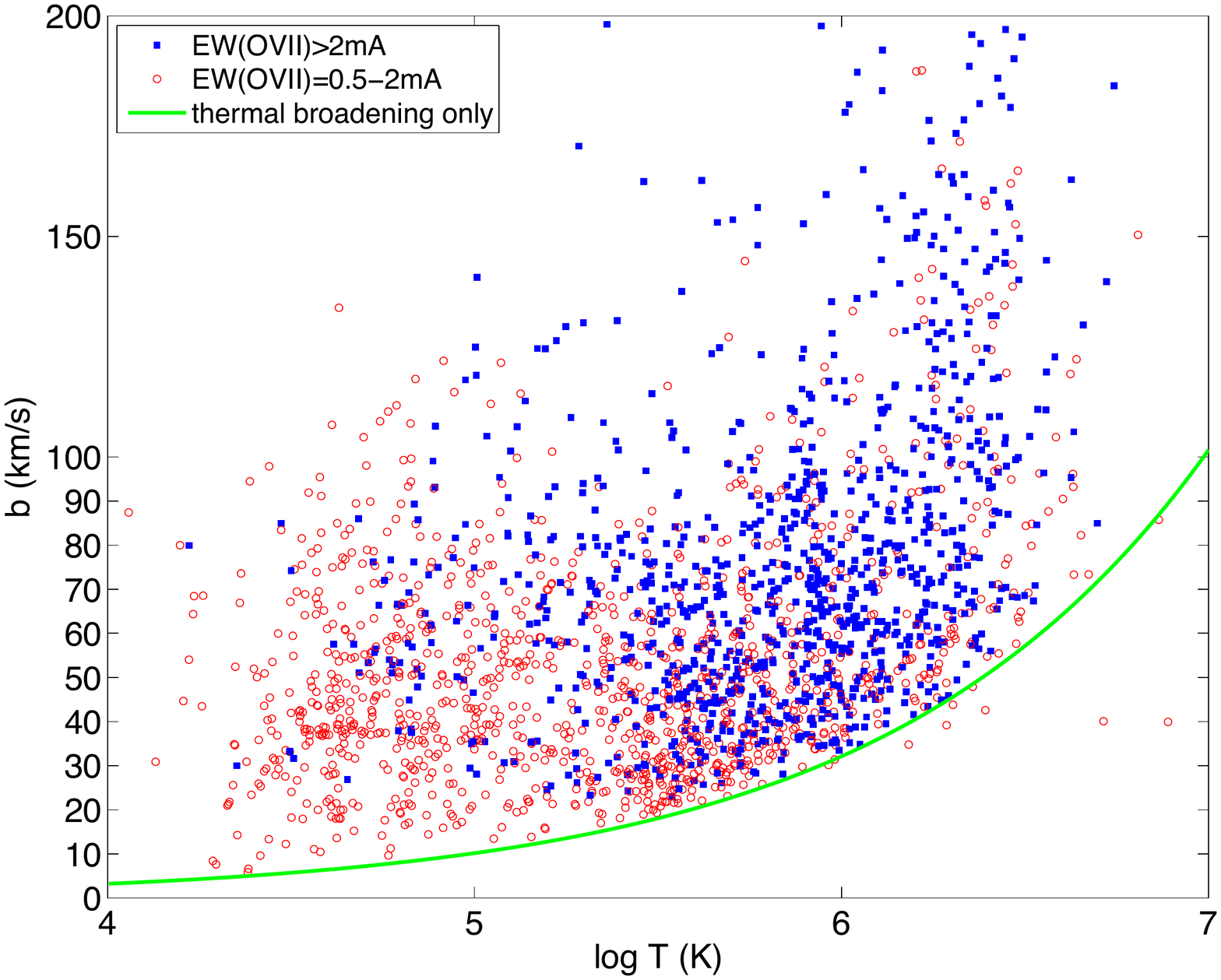}}
\vskip -0.4cm
\caption{
shows absorbers in the $b-T$ plane 
for O~VI line (top panel) and O~VII line (bottom panel).
Within each panel, we have broken up the absorbers into
strong ones (blue squares) and weak ones (red circles).
Only thermally broadened lines should follow the indicated solid green curve
(Eq. \ref{eq:bTeq}).
Also shown as right-pointing triangles are observed data of \citet[][]{2008Tripp} 
based on a joint analysis of $\lya$ and O~VI lines;
the location of each triangle is the best estimate of the temperature
and the rightmost tip of the attached line to each triangle represents
a $3\sigma$ upper limit.
}
\label{fig:bT_O6}
\end{figure}

\begin{figure}[H]
\centering
\resizebox{5.0in}{!}{\includegraphics[angle=0]{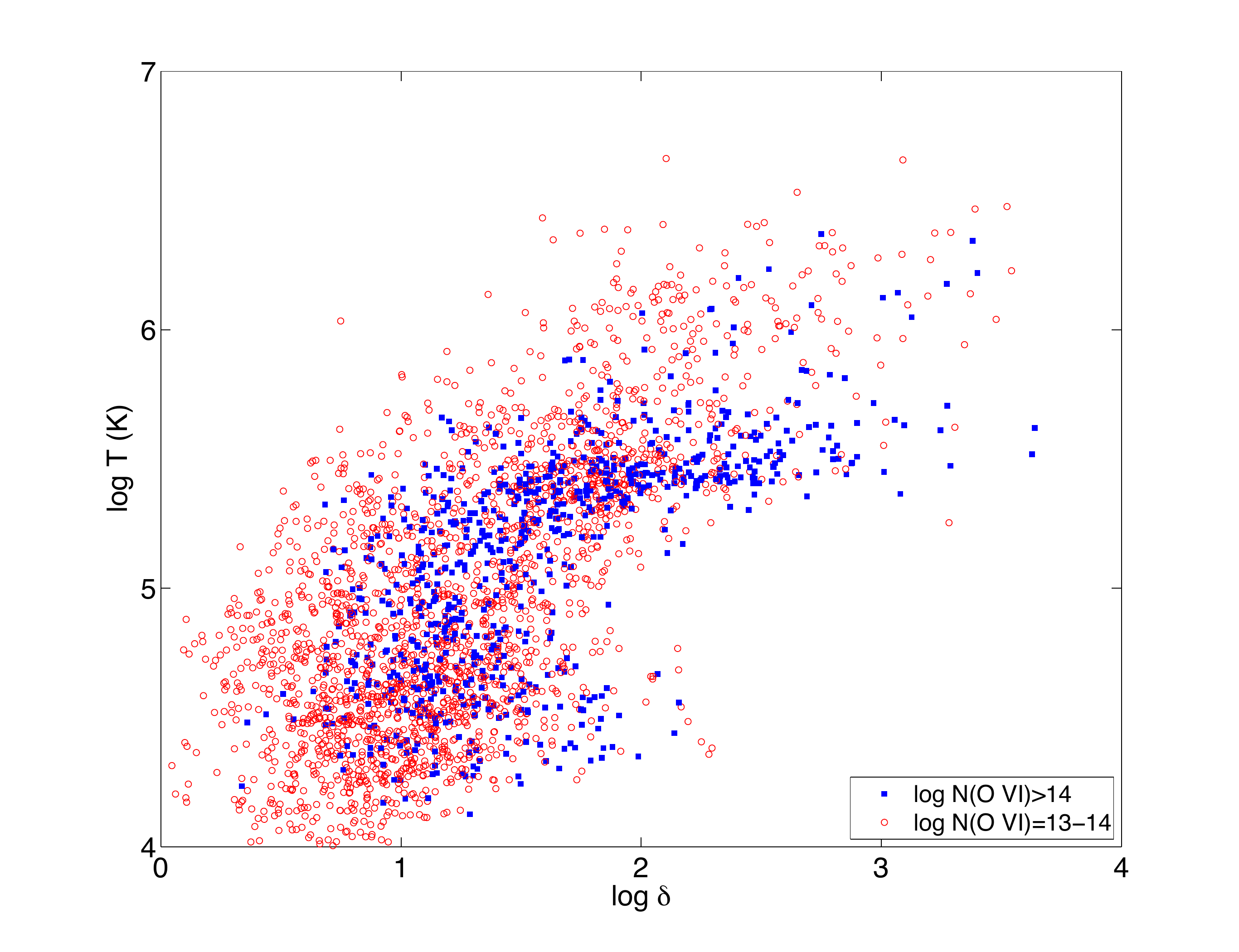}}
\vskip -1.0cm
\resizebox{5.0in}{!}{\includegraphics[angle=0]{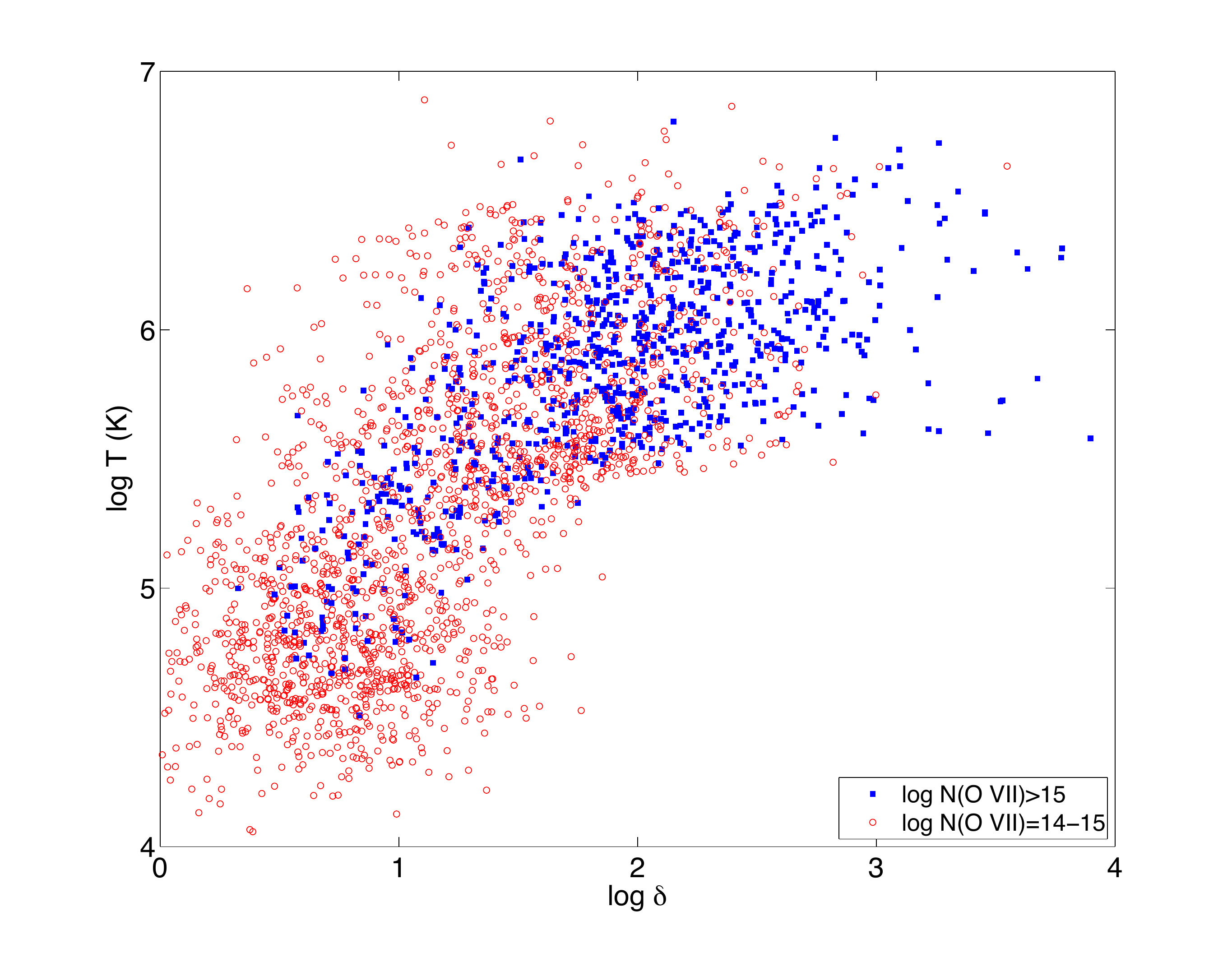}}
\vskip -0.4cm
\caption{
shows absorbers in the temperature-density plane 
for O~VI line (top panel) and O~VII line (bottom panel).
Within each panel, we have broken up the absorbers into
strong ones (blue squares) and weak ones (red circles).
}
\label{fig:Toverd_O6}
\end{figure}

\subsection{Physical Properties of O~VI and O~VII Lines}

In this subsection we will present physical properties of both O~VI and O~VII absorbers
and relationships between them.
For most of the figures below 
we will show results in pairs, one for O~VI and the other for O~VII, to facilitate comparisons.

Figure~\ref{fig:bT_O6} shows absorbers in the $b-{\rm T}$ plane 
for O~VI (top panel) and O~VII absorbers (bottom panel).
For thermal broadening only absorbers the $b-T$ relation would follow the solid green curve obeying this formula:
\begin{eqnarray}
\label{eq:bTeq}
{\rm b(O) = 10.16 (T/10^5K)^{1/2}}\kms. 
\end{eqnarray}
It is abundantly clear from Figure~\ref{fig:bT_O6} 
that $b$ is a poor indicator of absorber gas temperature.
Bulk velocity structures within each absorbing line are important.
For O~VI lines of equivalent width greater than $100$mA, it appears that 
bulk velocity structures are dominant over thermal broadening at all temperatures.
No line is seen to lie below the green curve, as expected.
All of the observationally derived temperature limits shown, based on a joint analysis of 
line profiles of well-matched coincidental $\lya$ and O~VI lines by 
\citet[][]{2008Tripp}, are seen to be fully consistent with our simulation results.
It is noted that velocity structures in unvirialized regions 
typically do not have gaussian distributions (in 1-d).
Caustic-like velocity structures are frequently seen that are reminiscent of
structure collapse along one dimension (e.g., Zeldovich pancake or filaments);
for anecdotal evidence see Figure~\ref{fig:spectra}.
Thus, we caution that temperatures derived
on the grounds of gaussian velocity profile \citep[e.g.,][]{2008Tripp} may be uncertain.
A more detailed analysis will be performed elsewhere.
The situations with respect to O~VII absorbers are similar to O~VI absorbers.

Figure~\ref{fig:Toverd_O6} 
shows absorbers in the temperature-density plane 
for O~VI (top panel) and O~VII absorbers (bottom panel).
In the top panel we see that strong O~VI absorbers with N(OVI)$\ge 10^{14}$cm$^{-2}$ 
have a large concentration at ($\delta$, ${\rm T}$)$=(10-300, \sim 10^{5.5}{\rm K})$
that corresponds to collisional ionization dominated O~VI population,
consistent with Figure~\ref{fig:dndz_O6}.
For weaker absorbers with N(OVI)$=10^{13-14}$cm$^{-2}$ we see that those with temperature above 
and those below $10^5$K are roughly equal, consistent with Figure~\ref{fig:dndz_O6};
the density distributions for the two subsets are rather different:
for the lower-temperature (T$<10^5$K) subset the gas density is concentrated around $\delta\sim 10$ 
that is photoionization dominated,
whereas for the higher-temperature (T$>10^5$K) subset the gas density 
is substantially spread out over $\delta \sim 3-3000$, which are mostly collisional ionization dominated.
Finally, we note that still weaker lines with N(OVI)$<10^{13}$cm$^{-2}$, not shown here,
are mostly photoionization dominated, as indicated in Figure~\ref{fig:dndz_O6}.
In the bottom panel we see that strong O~VII absorbers with N(OVI)$\ge 10^{15}$cm$^{-2}$ 
are predominantly collisionally ionized at ${\rm T}\sim 10^{5.5}-10^{6.5}$K and
$\delta\sim 10-1000$, with a small fraction of lines concentrated at an overdensity of $\sim 3-20$ and temperatures below $10^{5.5}$K.
For weaker absorbers with N(OVII)$=10^{14-15}$cm$^{-2}$ 
collisionally ionized ones at temperatures greater than $10^{5.5}$K and those photoionized
at lower temperatures are roughly 
 
\begin{figure}[H]
\centering
\resizebox{5.0in}{!}{\includegraphics[angle=0]{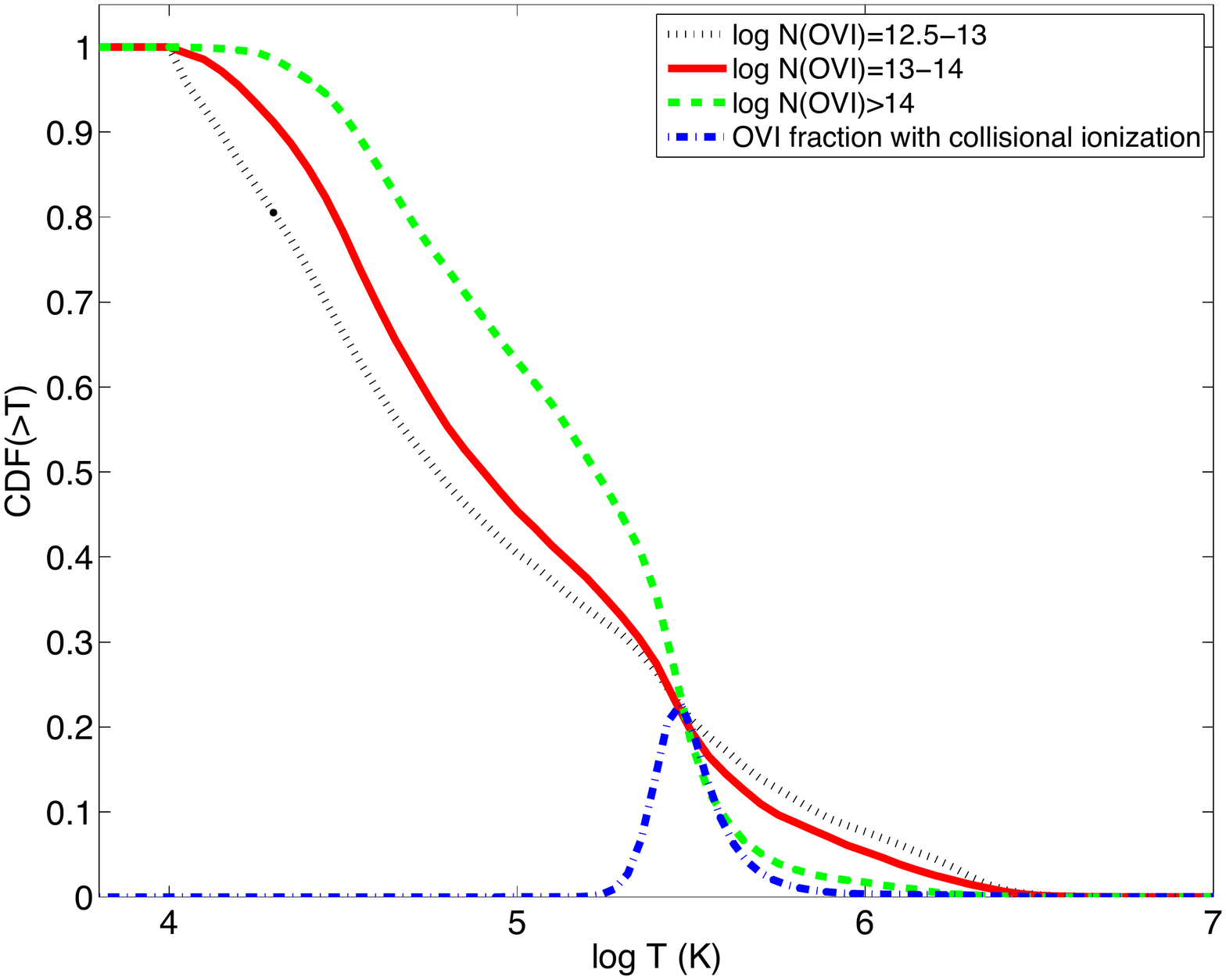}}
\vskip -1.0cm
\resizebox{5.0in}{!}{\includegraphics[angle=0]{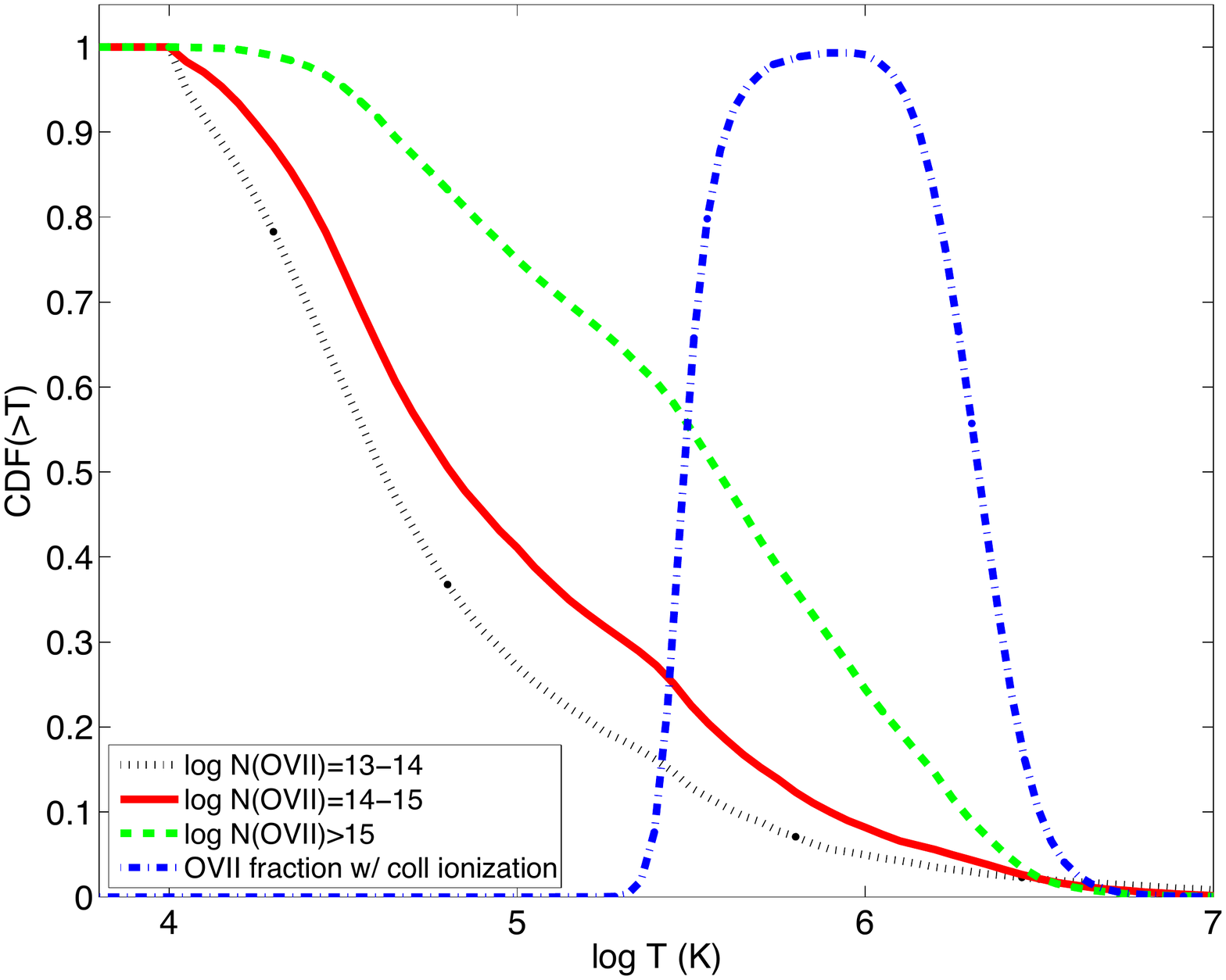}}
\vskip -0.4cm
\caption{
shows the cumulative probability distribution functions as a function of absorber temperature 
for three subsets of O~VI line with column densities
of $\log{\rm N(OVI) cm^2}=(12.5-13,13-14,>14)$ (top panel) 
and O~VII line of $\log{\rm N(OVI) cm^{2}}=(13-14,14-15,>15)$ (bottom panel).
The blue dot-dashed curve in the top (bottom) panel shows the O~VI (O~VII) fraction
as a function of gas temperature in the absence of photoionization.
}
\label{fig:This_O6}
\end{figure}
\noindent
comparable in numbers, 
consistent with Figure~\ref{fig:dndz_O7} below.
Our results for O~VI lines is broadly consistent with \citep[][]{2011Shull}.
\citet[][]{2011Shull} find a bimodal distribution of O~VI absorbers, 
one concentrating at 
($\delta$, ${\rm T}$)=($\sim 10$, $10^{4.5}$K) 
and the other at 
($\delta$, ${\rm T}$)=($\sim 100$, $10^{5.5}$K) (see their Figures 4,5).

Figure~\ref{fig:This_O6} 
shows the cumulative probability distribution functions as a function of absorber temperature 
for three subsets of O~VI lines with column densities
of $\log{\rm N(OVI)cm^2}=(12.5-13,13-14,>14)$ (top panel) 
and O~VII line of $\log{\rm N(OVI)cm^2}=(13-14,14-15,>15)$ (bottom panel).
We find that $(39\%, 43\%, 61\%)$ of O~VI absorbers in the 
column density ranges of $\log{\rm N(OVI)cm^2}=(12.5-13,13-14,>14)$ have temperature 
greater than $10^5$K,
$(25\%, 39\%, 73\%)$ of O~VII absorbers in the 
column density ranges of $\log{\rm N(OVI)cm^2}=(13-14,14-15,>15)$ have temperature greater than $10^5$K.
Our findings are in broad agreement with previous results obtained by our group 
\citep[e.g.,][]{2001Cen, 2006Cen, 2011Cen} and some other groups \citep[e.g.,][]{2011TepperGarcia, 2011Shull},
but in substantial disaccord with results of 
\citet[][]{2009Oppenheimer} and \citet[][]{2012Oppenheimer} 
who find that photo-ionized O~VI lines with temperature lower than $10^5$K 
make up the vast majority of O~VI lines across the column density range $\log{\rm N(OVI)cm^2}=12.5-15$.
Given the differences in simulation codes and in treatment of feedback processes,
we cannot completely ascertain the exact cause for the different results.
\noindent Nevertheless, the explanation given by \citet[][]{2012Oppenheimer} 
that the lack of metal mixing in their SPH simulations 
plays an important role in contributing to the difference is further elaborated here.

\citet[][]{2012Oppenheimer} find a large fraction of metal-carrying feedback SPH particles 
wound up in low density regions that have relatively high metallicity ($\sim 1\zsun$)
and low temperature (${\rm T}\sim 10^4$K).
As a result, they find 
low-density, high-metallicity and low-temperature photo-ionized O~VI absorbers
to dominate the overall O~VI absorber population in their SPH simulations.  
According to \citet[][]{2011TepperGarcia}, they repeat simulations with the same feedback model used in 
\citet[][]{2009Oppenheimer} and \citet[][]{2012Oppenheimer} 
but with metal heating included,
and are unable to reproduce the dominance of low-temperature photoionized O~VI absorbers seen in the latter.
This leads them to conclude that lack of metal heating, in the presence of high-metallicity feedback SPH particles, 
is the cause of the dominance of low-density, high-metallicity and low-temperature photo-ionized O~VI absorbers
found in \citet[][]{2009Oppenheimer} and \citet[][]{2012Oppenheimer}.
We suggest that this overcooling problem may have been exacerbated by lack of metal mixing. 
Consistent with this conjecture, while \citet[][]{2011TepperGarcia} suffer less severely from the metal overcooling problem
(because of metal heating),
the median metallicity of their O~VI absorbers is still $\sim 0.6\zsun$, substantially higher than that of our O~VI absorbers, 
$Z\sim 0.03-0.3\zsun$,
even though their overall abundance of O~VI absorbers is lower than observed by a factor of $\sim 2$.
This noticeable difference in metallicity may be rooted in lack of metal mixing in theirs.

\noindent 
As we will show later (see Figure~\ref{fig:ZN} below), 
the metallicity of simulated O~VI in our simulations appears to better match observations.
Despite that, it is desirable to directly probe the physical nature of O~VI absorbers to test models 
by different simulation groups.
One major difference between SPH simulations 
\citep[e.g.,][]{2011TepperGarcia, 2012Oppenheimer}
and AMR simulations \citep[e.g.,][and this work]{2011Smith} is that
the former predict metallicity distributions that are peaked at $(0.6-1)\zsun$
compared to peaks of $\sim (0.05-0.2)\zsun$ in the latter.
In addition, in the latter positive correlations between metallicity and O~VI column density
and between metallicity and temperature are expected, whereas in the former the opposite or little correlation
seems to be true.
Therefore, direct measurements of O~VI metallicity and correlations between metallicity and
other physical quantities would provide a good discriminator.
Putting differences between SPH simulations of WHIM by different groups aside,
what is in common among them is the dominance 
of low-density ($\delta \le 100$) O~VI absorbers at all column densities.
In the AMR simulations it is found that the collisionally ionized O~VI absorbers,
with density broadly peaked at $\delta\sim 100$,
dominate (by 2 to 1) over photoionized O~VI absorbers 
for ${\rm N(OV)\ge 10^{14}}$cm$^{-2}$ population. 
Given these significant differences between SPH and AMR simulations,
we suggest a new test, namely, the cross-correlation function between galaxies and strong 
[${\rm N(O~VI)\ge 10^{14}}$cm$^{-2}$] O~VI absorbers.
Available observations appear to point to strong correlations at relatively small scales $\le 300-700$kpc
between luminous galaxies ($\ge 0.1L_*$) and ${\rm N(OV)\ge 10^{13.2-13.8}}$cm$^{-2}$ O~VI absorbers
at $z=0-0.5$ \citep[e.g.,][]{2006Stocke, 2009Chen,2011Prochaska}. 
We expect AMR simulated O~VI absorbers to show stronger small-scale cross-correlations with galaxies
than SPH simulated O~VI absorbers thanks to the predicted dominance of O~VI absorbers in low density
(but higher metallicity) regions in the latter that are at larger distances from galaxies.
When an adequate observational sample of Ne~VIII absorbers becomes available,
galaxy-Ne~VIII may provide a still more sensitive test between photoionization models suggested by some SPH simulations
and AMR simulations, because the N~VIII line needs a still higher 
temperature to collisionally ionize and hence the contrast is still higher between the simulations.
However, the shorter wavelengths of the Ne VIII lines at 770\AA\  and 780\AA\  require galaxies at $z>0.47$ to shift into the HST (COS or STIS) observable band,   
for which current galaxy surveys will only be able to probe most luminous galaxies ($L>L_*$).
Detailed calculations and comparisons to observations will be needed to ascertain these
expectations to nail down their physical nature and to constrain feedback models.


The rapid rise in the cummulative fraction in the temperature range $\log {\rm T}=5.5\pm 0.1$
in the top panel of Figure~\ref{fig:This_O6} 
reflects the concentration of collisionally ionized O~VI lines in that temperature range,
supported the blue dot-dashed curve showing the collisional ionization fraction of O~VI as a function of temperature.
This feature is most prominent in the high column density O~VI population with 
$\log{\rm N(OVI) cm^2}>14$ (green dashed curve), simply stating the fact that collisional ionization
is dominant in high column density O~VI lines.
For O~VII there is a similar feature except that it is substantially broader at 
$\log {\rm T}=6.0\pm 0.4$, which is consistent with the 
ionization fraction of O~VII as a function of temperature in the collisional ionization dominated regime,
shown as the blue dot-dashed curve in the bottom panel of Figure~\ref{fig:This_O6}.
For O~VI lines with column density in the range
$\log{\rm N(OVI) cm^2}=12.5-14$ we see a relative dearth of absorbers in the temperature range
$\log {\rm T}=4.8-5.4$, a regime where neither collisional ionization nor photoionization
is effective due to structured multi-phase medium (i.e., positive correlation between density and temperature
in this regime, see Figure~\ref{fig:phase} below);
at still lower temperature $\log {\rm T}<4.8$ (and low density due to the positive correlation),
the curve displays a rapid ascent due to its entry into the photoionization dominated regime.
Analogous behaviors and explanations can be said for O~VII absorbers.

\begin{figure}[H]
\centering
\resizebox{5.0in}{!}{\includegraphics[angle=0]{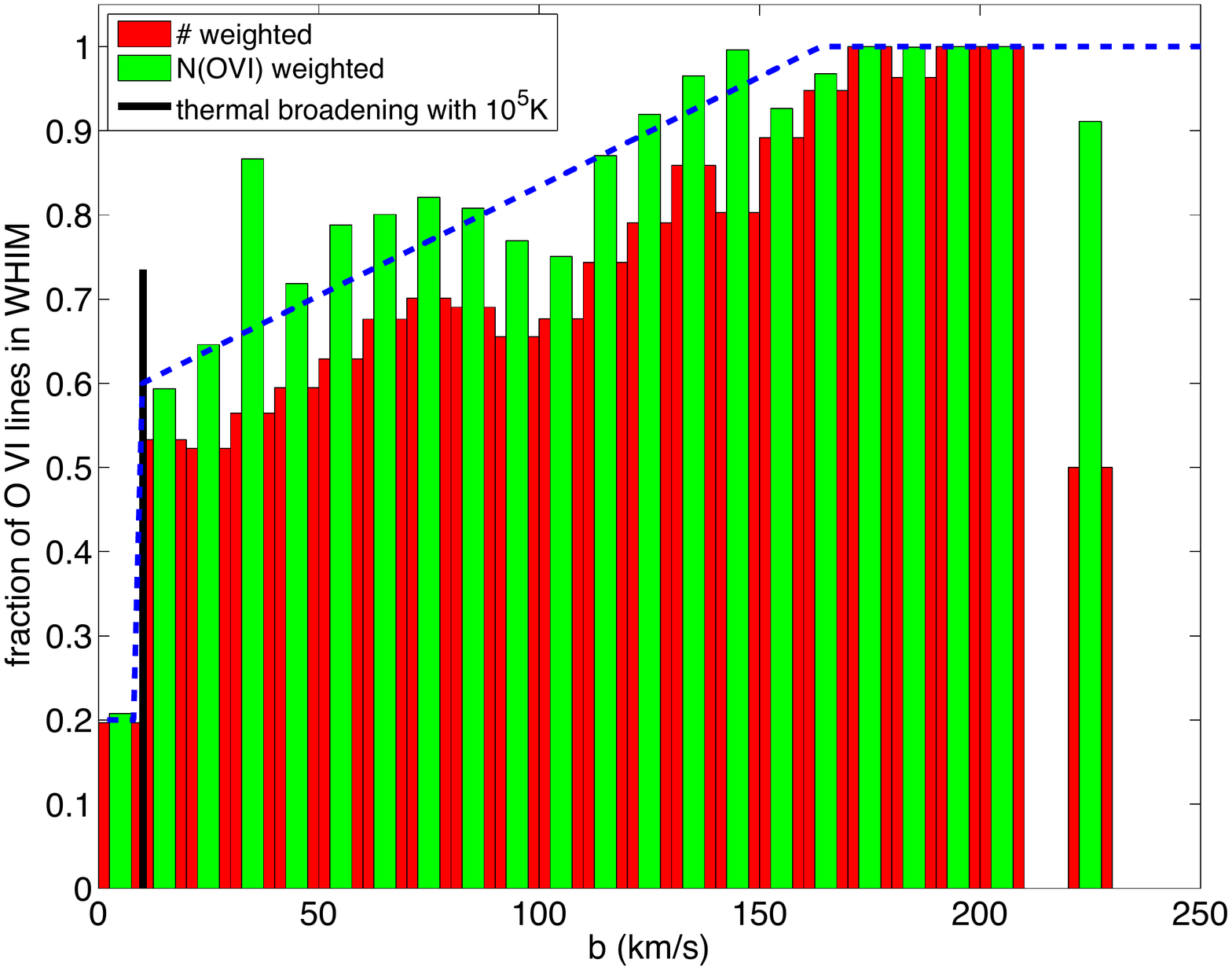}}
\vskip -1.0cm
\resizebox{5.0in}{!}{\includegraphics[angle=0]{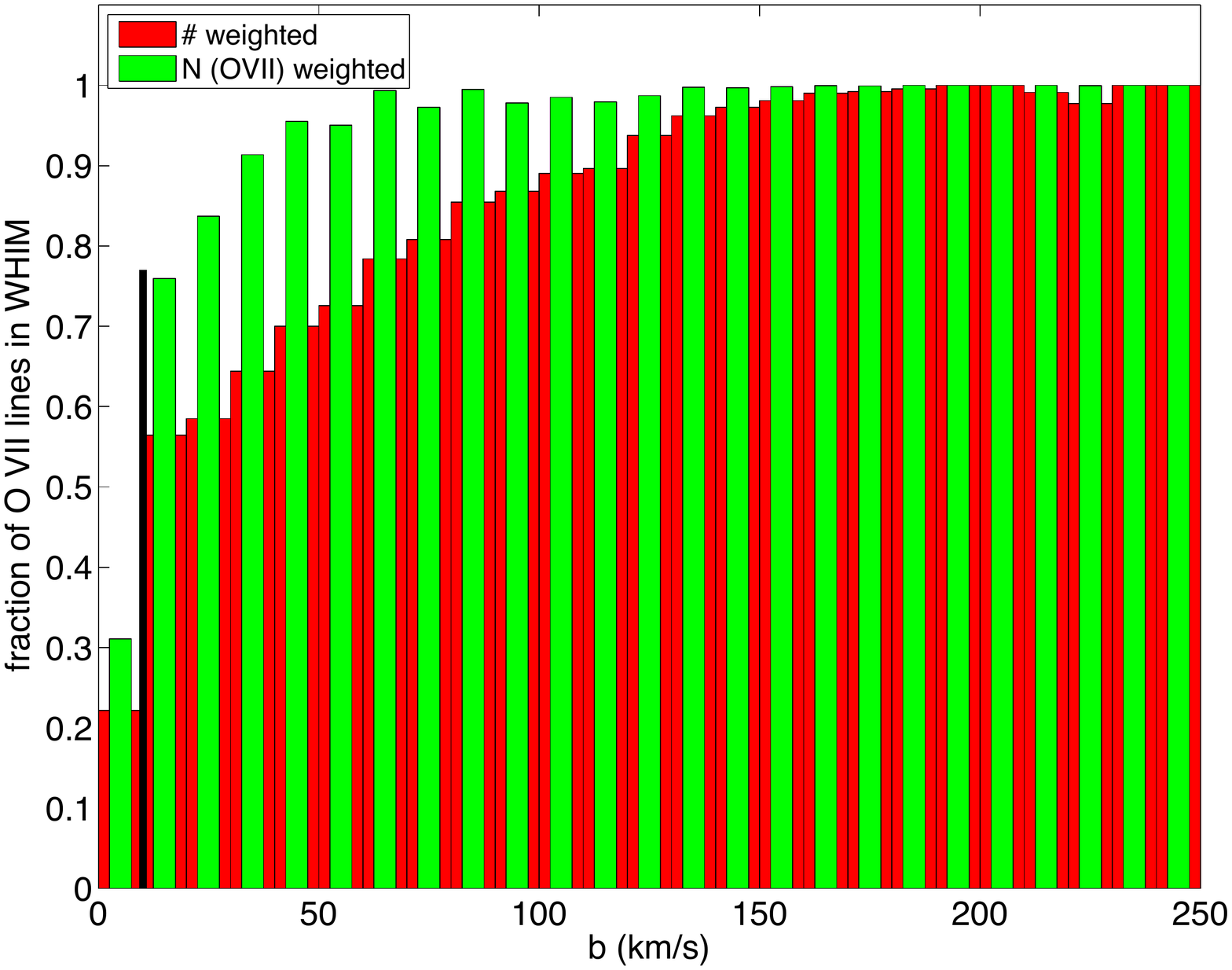}}
\vskip -0.4cm
\caption{
shows the fraction of O~VI (top panel) and O~VII (bottom panel) absorbers
that is in WHIM temperature range of $10^5-10^7$K as a function of $b$.
The red and green histograms are number and column density weighted, respectively,
including only lines with column density above $10^{13}$cm$^{-2}$ in the case of O~VI and $10^{14}$cm$^{-2}$ for O~VII.
The vertical black line indicates $b$ for a purely thermally broadened line at a temperature of $10^5$K.
The approximate fitting curve indicated by blue dashed line is given in Equation (2).
}
\label{fig:O6WHIMfrac}
\end{figure}

\begin{figure}[H]
\centering
\resizebox{5.0in}{!}{\includegraphics[angle=0]{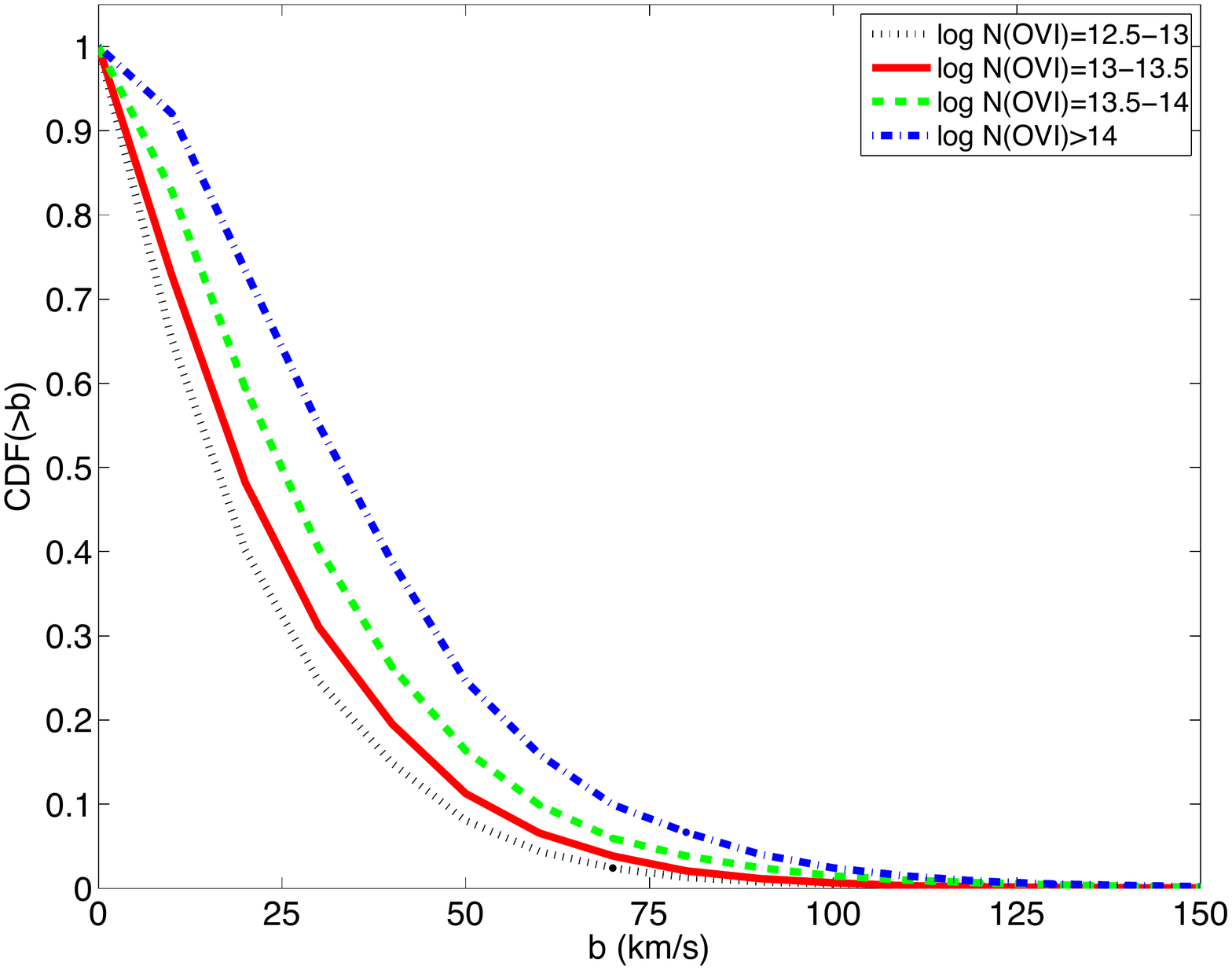}}
\vskip -1.0cm
\resizebox{5.0in}{!}{\includegraphics[angle=0]{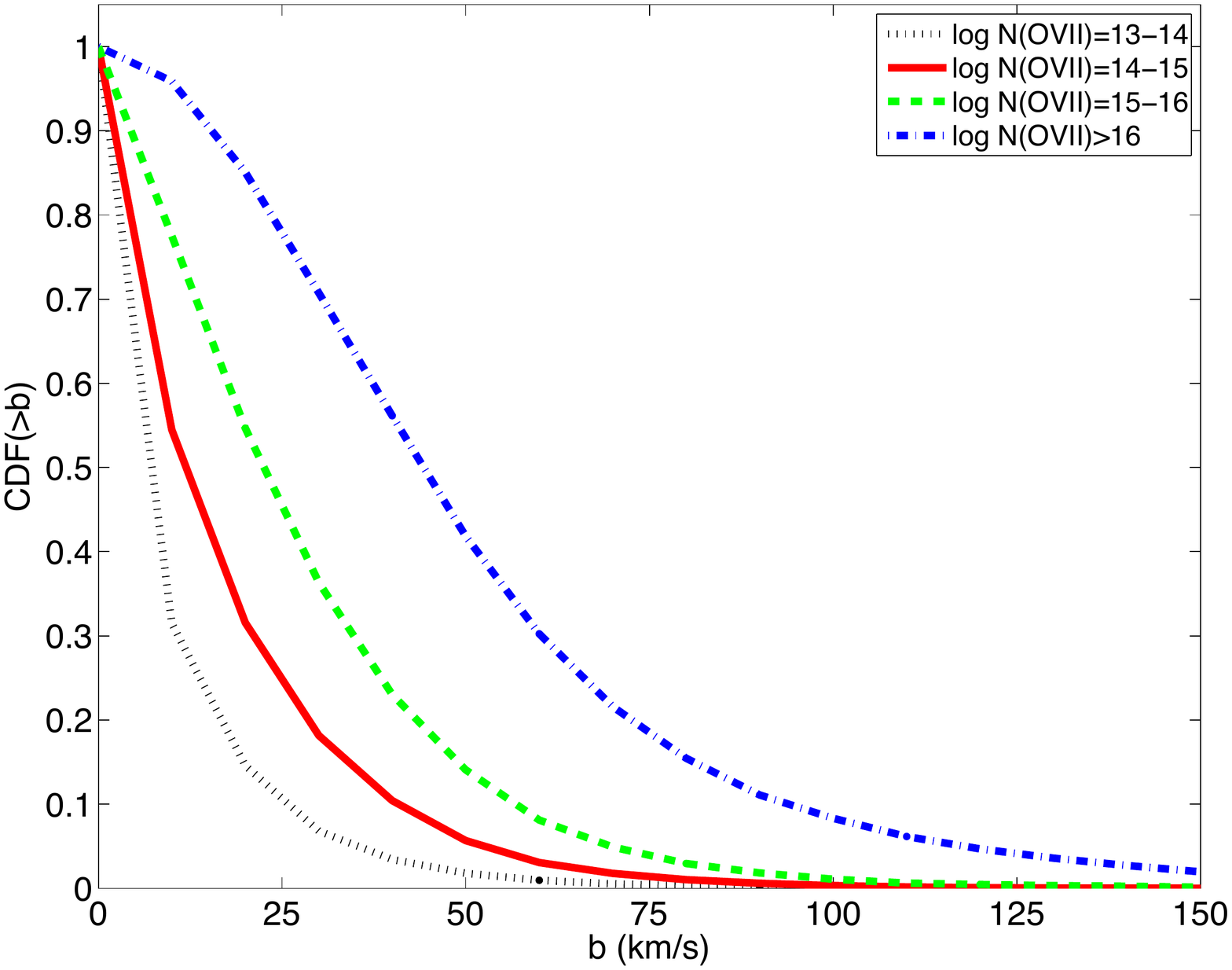}}
\vskip -0.4cm
\caption{
Top panel shows four cumulative probability distribution functions as a function of $b$ 
for four subsets of O~VI lines in the four column density ranges:
$\log{\rm N(OVI)cm^2}=12.5-13$ (black dotted curve),
$\log{\rm N(OVI)cm^2}=13-13.5$ (red solid curve),
$\log{\rm N(OVI)cm^2}=13.5-14$ (green dashed curve) and 
$\log{\rm N(OVI)cm^2}>14$ (blue dot-dashed curve).
Bottom panel shows four cumulative probability distribution functions as a function of $b$ 
for four subsets of O~VII lines in the four column density ranges:
$\log{\rm N(OVI)cm^2}=13-14$ (black dotted curve),
$\log{\rm N(OVI)cm^2}=14-15$ (red solid curve),
$\log{\rm N(OVI)cm^2}=15-16$ (green dashed curve) and 
$\log{\rm N(OVI)cm^2}>16$ (blue dot-dashed curve).
}
\label{fig:bhis}
\end{figure}

We see earlier in Figure~\ref{fig:bT_O6} that 
$b$ is not a good indicator of the temperature of absorbing gas.
It is thus useful to quantify the fraction of absorbers at a given $b$ whose temperature is in the WHIM regime.
Figure~\ref{fig:O6WHIMfrac} shows
the fraction of O~VI (top panel) and O~VII (bottom panel) absorbers
that is in WHIM temperature range of $10^5-10^7$K as a function of $b$.
Broadly speaking, above the threshold (thermally broadened Doppler width of $10.16$km/s for a gas at temperature of $10^5$K),
the WHIM fraction is dominant at $\ge 50\%$ for both O~VI and O~VII lines,
but only close to $100\%$ when $b$ is well in excess of $100$km/s.
This again indicates the origin of the O~VI absorbing gas whose random motions 
are far from completely thermalized, consistent with its (mostly) intergalactic nature.
The approximate fitting curve (blue curve) for the column density weighted hisotogram for O~VI shown in
Figure~\ref{fig:O6WHIMfrac} can be formulated as the following equation:
\begin{eqnarray}
\label{eq:O6eq}
{\rm f(OVI)} &=& 0.20 \hfill\quad\quad{\rm for}\quad\quad b<10\kms  \nonumber \\
&=& 0.0026(b-10) + 0.6  \hfill\quad\quad{\rm for}\quad\quad b=10-160\kms  \nonumber \\
&=& 1 \hfill\quad\quad{\rm for}\quad\quad b>160\kms
\end{eqnarray}
As already indicated in Figure~\ref{fig:bN_O6} that a substantial fraction of broad but shallow absorbers 
may be missing in current observational data, here we quantify it further.
Figure~\ref{fig:bhis} shows four cumulative probability distribution functions as a function of $b$ 
for four subsets of O~VI (top panel) and O~VII (bottom panel) lines of differing column densities.
To give some quantitative numbers, 
(15\%, 20\%, 26\%, 39\%) of O~VI absorbers with 
$\log{\rm N(O~VI)}=(12.5-13, 13-13.5, 13.5-14, >14)$ have $b>40\kms$;
the fractions drop to (1\%, 2\%, 4\%, 7\%) for $b>80\kms$.
Similarly, (17\%, 46\%, 77\%, 88\%) of O~VII absorbers with 
$\log{\rm N(O~VI)}=(13-14, 14-15, 15-16, >16)$ have $b>40\kms$,
with (2\%, 9\%, 29\%, 30\%) having $b>80\kms$.
With COS observations of substantially higher sensitivities,
broad O~VI lines are begun to be detected \citep[][]{2010Savage}.
A direct, statistical comparison between simulation 

\noindent
results found here and observations
will be possible in the near future. 
It will be extremely interesting to see
if there is indeed a large population of broad but shallow O~VI lines still missing.
That is also important, because, if that is verified,
one will have more confidence on the results for O~VII lines,
which suggest that the O~VII lines may be substantially broader than
a typical thermally broadened width of $40-50\kms$.
Additional useful information is properties of other lines,
including $\lya$, which we will present in a subsequent paper.
Since the expected $b$ of O~VII lines is still substantially smaller than
spectral resolution of Chandra and XMM-Newton X-ray instruments,
it does not make much difference for extant observations.
However, it should be taken into consideration in designing future X-ray telescopes to probe WHIM in absorption or emission
\citep[e.g.,][]{2012Yao}.

\begin{figure}[H]
\hskip -0.7cm
\centering
\resizebox{6.5in}{!}{\includegraphics[angle=0]{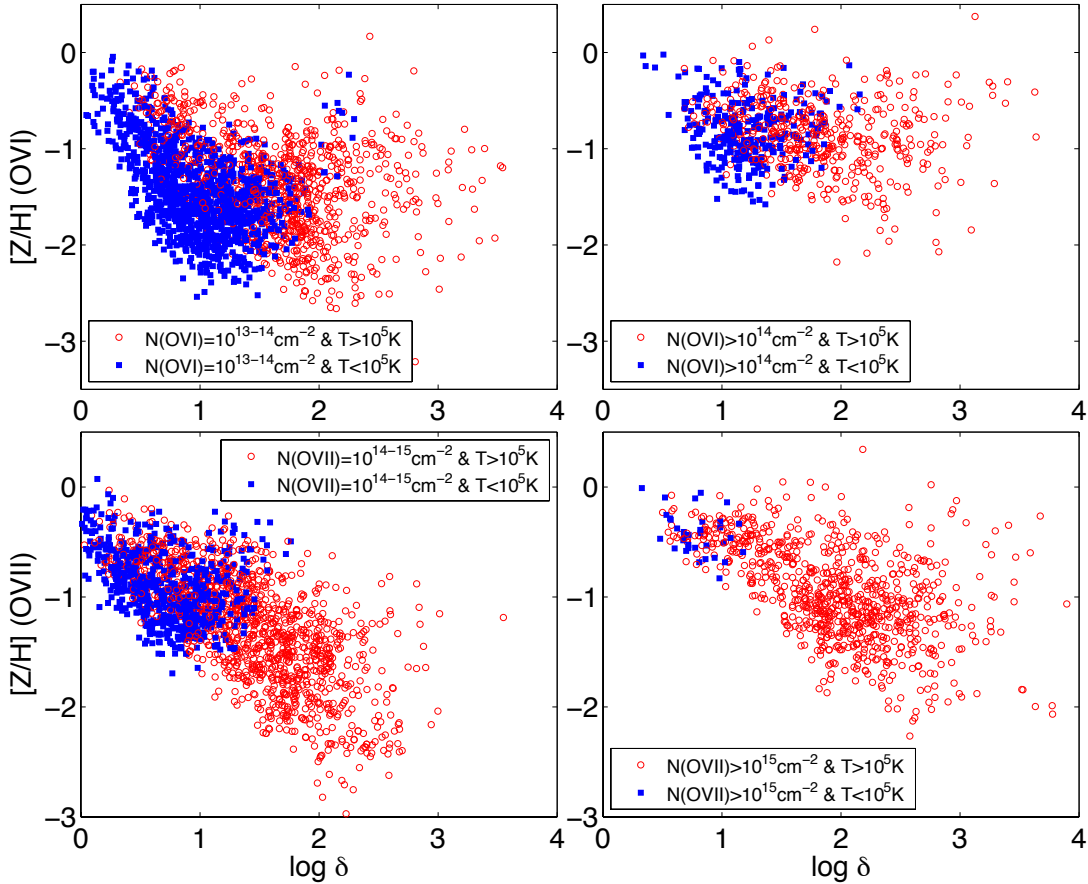}}
\vskip -0.4cm
\caption{
shows absorbers in the metallicity-overdensity plane
for O~VI line with
${\rm N(OVI)}=10^{13-14}$cm$^{-2}$
(top left panel)
and ${\rm N(OVI)}>10^{14}$cm$^{-2}$ (top right panel.
The bottom two panels show
O~OVII line with
${\rm N(OVII)}=10^{14-15}$cm$^{-2}$
(bottom left panel)
and
${\rm N(OVII)}>10^{15}$cm$^{-2}$ (bottom right panel).
Within each panel, we have broken up the absorbers into
two subsets using temperature: $T>10^5$K (red circles)
and $T<10^5$K (blue squares).
}
\label{fig:Zoverd}
\end{figure}

Figure~\ref{fig:Zoverd} shows absorbers in the metallicity-overdensity plane.
The apparent anti-correlation between metallicity and overdensity with a log slope
of approximately $-1$ for the mostly collisional ionization dominated population
(red circles) is simply due to the fact that
the absorbers near the chosen column density cutoff 
dominate the numbers and that 
collisional ionization rate is density independent.
Similarly, the apparent anti-correlation between metallicity and overdensity with a log slope
of approximately $-2$ for the mostly photoionization dominated population
(blue squares) is due to the fact that photo-ionization fraction is proportional to density.
So one should not be misled to believe that there is an anti-correlation between
gas metallicity and overdensity in general; the opposite is in fact true (see Figure~\ref{fig:metal_overd} below).
In Figure~\ref{fig:O6Zb} we project two subsets of absorbers onto the metallicity-$b$ plane.
Because of complex behaviors seen in Figure~\ref{fig:Zoverd} and the additional role
played by complex temperature and velocity distributions, one may not be surprised to see
the large dispersions in metallicity at a given $b$.
The metallicity distribution is seen to be, to zero-order within the large dispersions, nearly independent of $b$.
When metallicity of O~VI and O~VII absorbers can be measured directly in the future, this prediction may be tested.
No further detailed information on this shall be given here due to its still more futuristic nature 
in terms of observability,
except noting that the weak trends can be understood and these trends are dependent upon the column density cuts.

Figure~\ref{fig:ZN} shows the mean metallicity as a function of column density 
for O~VI (red circles) and O~VII absorbers (blue squares).
We see that a substantial dispersion of about 0.5-1 dex is present for all column density bins.
The mean metallicity for O~VI lines increases by 0.9~dex from
${\rm [Z/H]}\sim -1.3$ at ${\rm N(OVI)}=10^{13}$cm$^{-2}$ to 
${\rm [Z/H]}\sim -0.4$ at ${\rm N(OVI)}=10^{15}$cm$^{-2}$.
For O~VII lines the mean metallicity increases by 0.4~dex from
${\rm [Z/H]}\sim -1.1$ at $10^{14}$cm$^{-2}$ to 
${\rm [Z/H]}\sim -0.7$ at $10^{16}$cm$^{-2}$.
The trend of increasing metallicity with increasing column density 
is consistent with the overall trend that higher density regions, on average,
have higher metallicity, at least in the density range of interest here (see Figure~\ref{fig:Zoverd} below).
It is noted that the mean metallicity for O~VII absorbers is, on average, lower than that for O~VI lines at
a fixed column density for the respective ions.
This and some other relative behaviors between O~VI and O~VII seen in Figure~\ref{fig:ZN} 
merely reflect the facts (1) that the 
product of oscillator strength and restframe wavelength of O~VII line
is about a factor of $10$ lower than that of O~VI, (2) the peak collisional ionization fraction  
for O~VII is about 
a factor of 5 
higher than that of O~VI, and (3) the peak width for collisional ionization 
temperature for O~VII is larger by a factor of $\sim 3$ than
that of O~VI (see Figure~\ref{fig:This_O6}).
Examination of the C+P (collisional + photoionization) model 
with distributed feedback (which is closest to our feedback model)
in Figure 17 of \citet[][]{2011Smith} reveals that the average metallicity
increases from $\sim -1.0$ to $\sim 0.0$ for ${\rm N_{OVI}}$ from 
$10^{12}$cm$^{-2}$ to $10^{15}$cm$^{-2}$, 
which should be compared to an increase of metallicity from $\sim -1.5$ to $\sim -0.5$.
Thus, our results are in good agreement with 
\citet[][]{2011Smith} except that their metallicity is uniformly higher by a factor of $\sim 3$.

While there are substantial disagreements among the SPH and AMR simulations with respect to 
the metallicity of O~VI absorbers,
it is fair to say that a median value of $0.1-1\zsun$ encompasses them. 
Given that, some quantitative physical considerations are useful here.
The cooling time for gas of $\delta=100$, $T=10^{5.5}$K and $Z=0.1\zsun$ at $z=0$ is $\sim 0.05t_H$
($t_H$ is the Hubble time at $z=0$) (this already takes into account metal heating by the X-ray background;
it should be noted that the X-ray background at $z\sim 0$ is still quite uncertain \citep[e.g.,][]{2011Shull}).
This indicates that the O~VI-bearing gas of $T\sim 10^{5.5}$K 
and $\delta \ge 100$, in the absence of other balancing heating processes, 
can only spend a small fraction of a Hubble time at the temperature for optimal O~VI production via collisional ionization.
This has two implications.
First, O~VI absorbers at $\delta \ge 100\times (Z/0.1\zsun)^{-1}$ is transient in nature
and their appearance requires either constant heating of colder gas or higher temperature gas cooling through.
Which process is more responsible for O~VI production will be investigated in a future study.
Second, the metal cooling that is linearly proportional to gas metallicity 
may give rise to an interesting ``selection effect", where
high metallicity O~VI gas in dense regions, having shorter cooling time than lower metallicity
O~VI gas of the same density, would preferentially remove itself
from being O~VI productive by cooling,
leaving behind only lower metallicity gas at O~VI-bearing temperatures. 
We suggest that this selection effect may have contributed to a much reduced proportion of
collisionally ionized O~VI lines in SPH simulations that lack adequate metal mixing;
in other words, dense metal ``bullets" of SPH particles
either cools very quickly to $\sim 10^{4}$K or they have reached regions of sufficiently low density before
that happens.
The results of \citet[][]{2012Oppenheimer} appear to suggest, 
in the context of this scenario, that the feedback metal-bearing SPH particles have cooled
to $\sim 10^{4}$K, before they can reach low density regions to avoid severe cooling,
thus resulting in high-metallicity, low-density, photoionized O~VI lines
when they eventually wind up in low density regions.
An analogous situation occurs in \citet[][]{2011TepperGarcia} SPH simulations
but with two significant differences from those of \citet[][]{2012Oppenheimer}:
(1) in the former the inclusion of metal heating (due to photoionization of metal species)
keeps the corresponding SPH particles at a higher temperature floor
($\sim 10^{4.5-5}$K barring adiabatic cooling) than in the latter, and
(2) ``smoothed" metallicity used in the former to compute metal cooling/heating rates
has reduced the metal cooling effects (which still dominate over metal heating at $T\ge 10^5$K) 
compared to the case without such smoothing in the latter.

\begin{figure}[ht]
\hskip -0.7cm
\centering
\resizebox{3.5in}{!}{\includegraphics[angle=0]{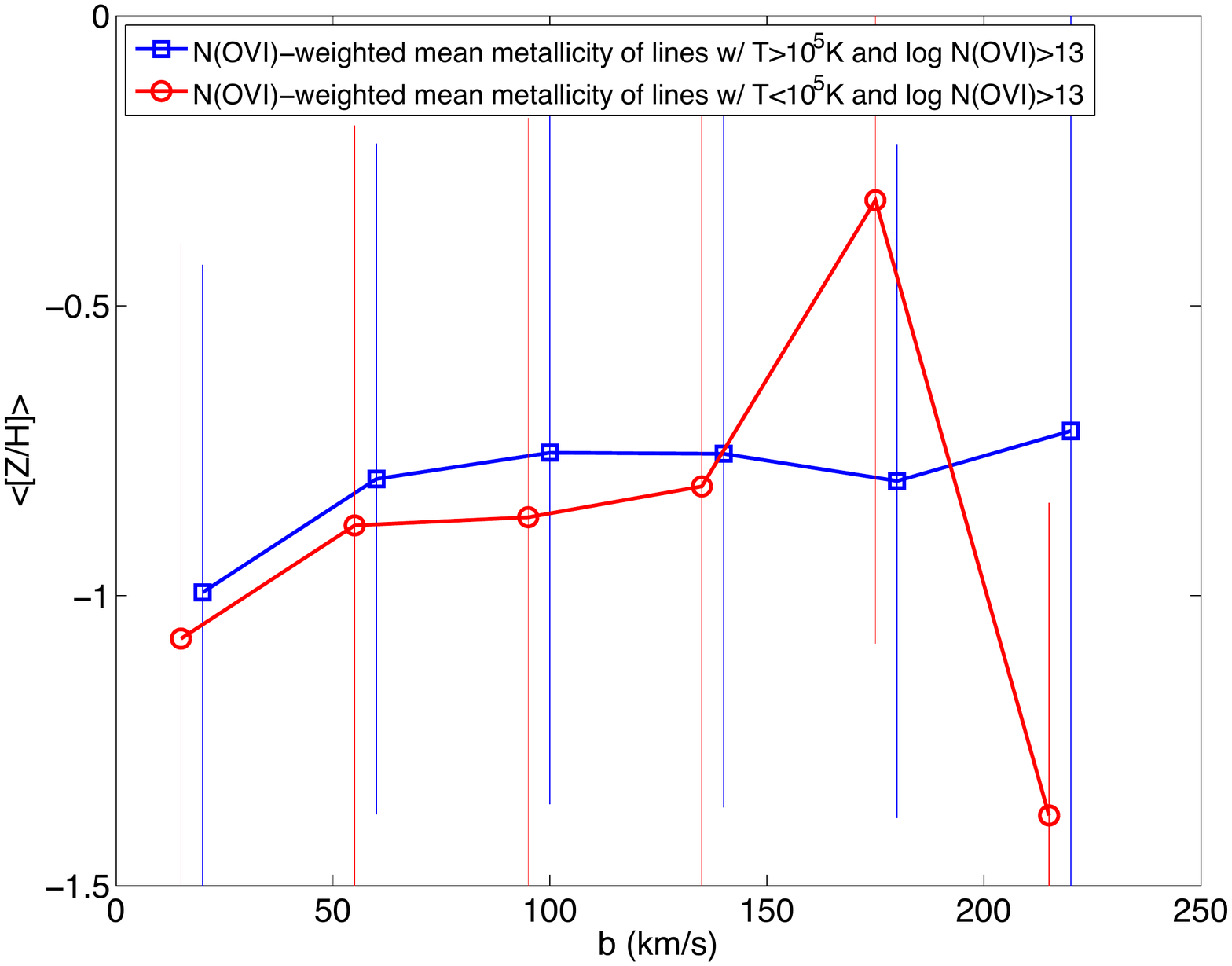}}
\hskip -0.5in
\resizebox{3.5in}{!}{\includegraphics[angle=0]{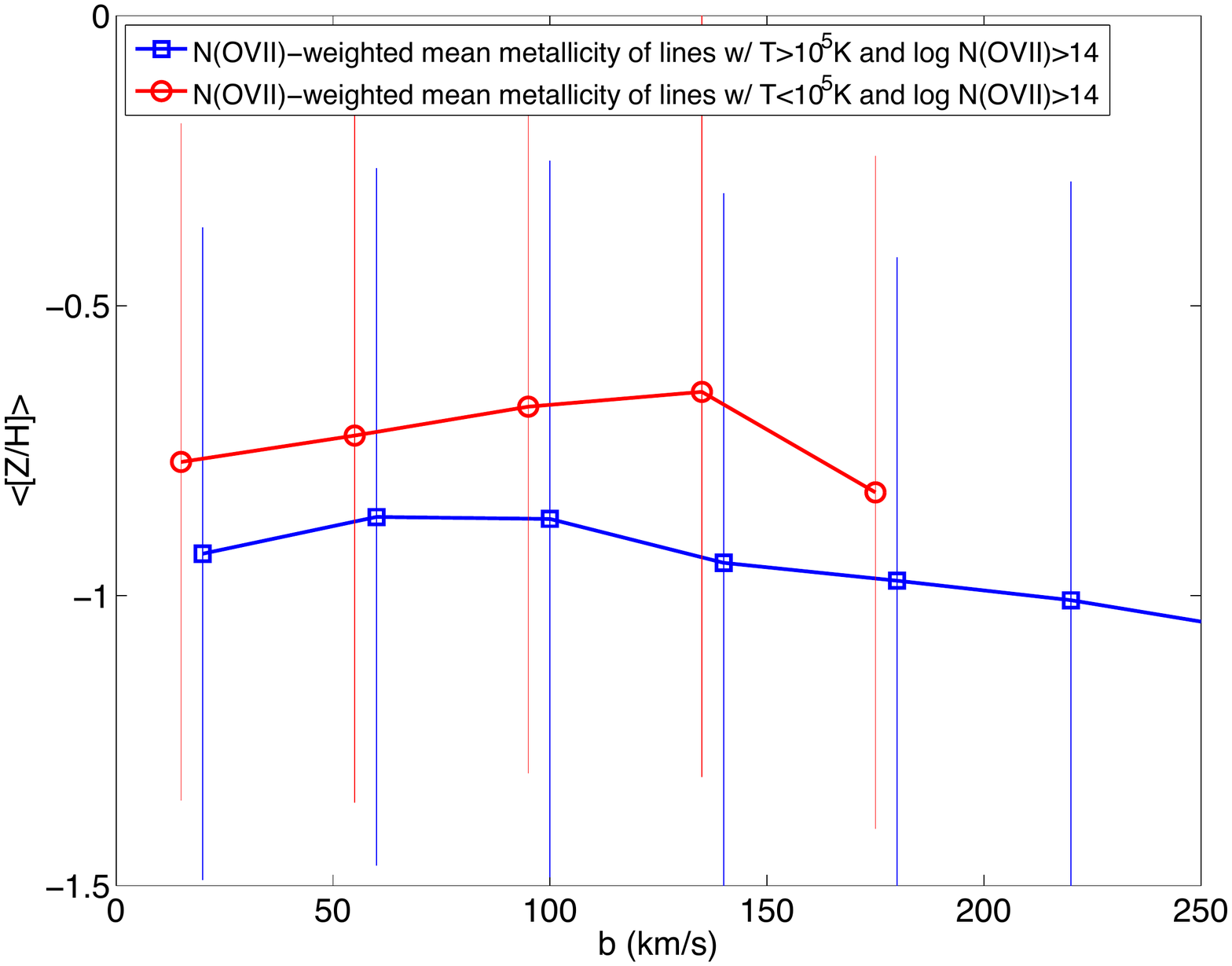}}
\vskip -0.4cm
\caption{
shows the mean absorber metallicity as a function of $b$ 
for O~VI line with column density above $10^{13}$cm$^{-2}$
(top panel) and O~VII line with column density above $10^{14}$cm$^{-2}$ (bottom panel).
Within each panel, we have broken up the absorbers into
two subsets using temperature: $T>10^5$K (blue squares) and $T<10^5$K (red circles).
}
\label{fig:O6Zb}
\end{figure}

\begin{figure}[h!]
\hskip -0.7cm
\centering
\resizebox{5.0in}{!}{\includegraphics[angle=0]{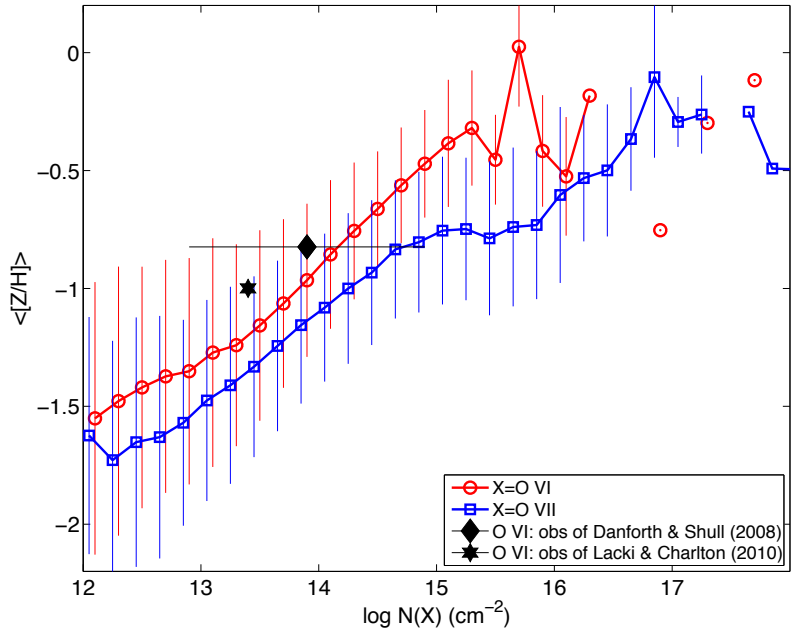}}
\vskip -0.5cm
\caption{
shows the mean metallicity as a function of column density 
for O~VI (red open circles) and O~VII (blue open squares) lines.
Also shown as solid symbols are observational data.
It is likely that the observational errorbars are underestimated.
}
\label{fig:ZN}
\end{figure}

\subsection{Coincidence Between O~VI and O~VII Lines}

\begin{figure}[ht]
\hskip -0.7cm
\centering
\resizebox{6.0in}{!}{\includegraphics[angle=0]{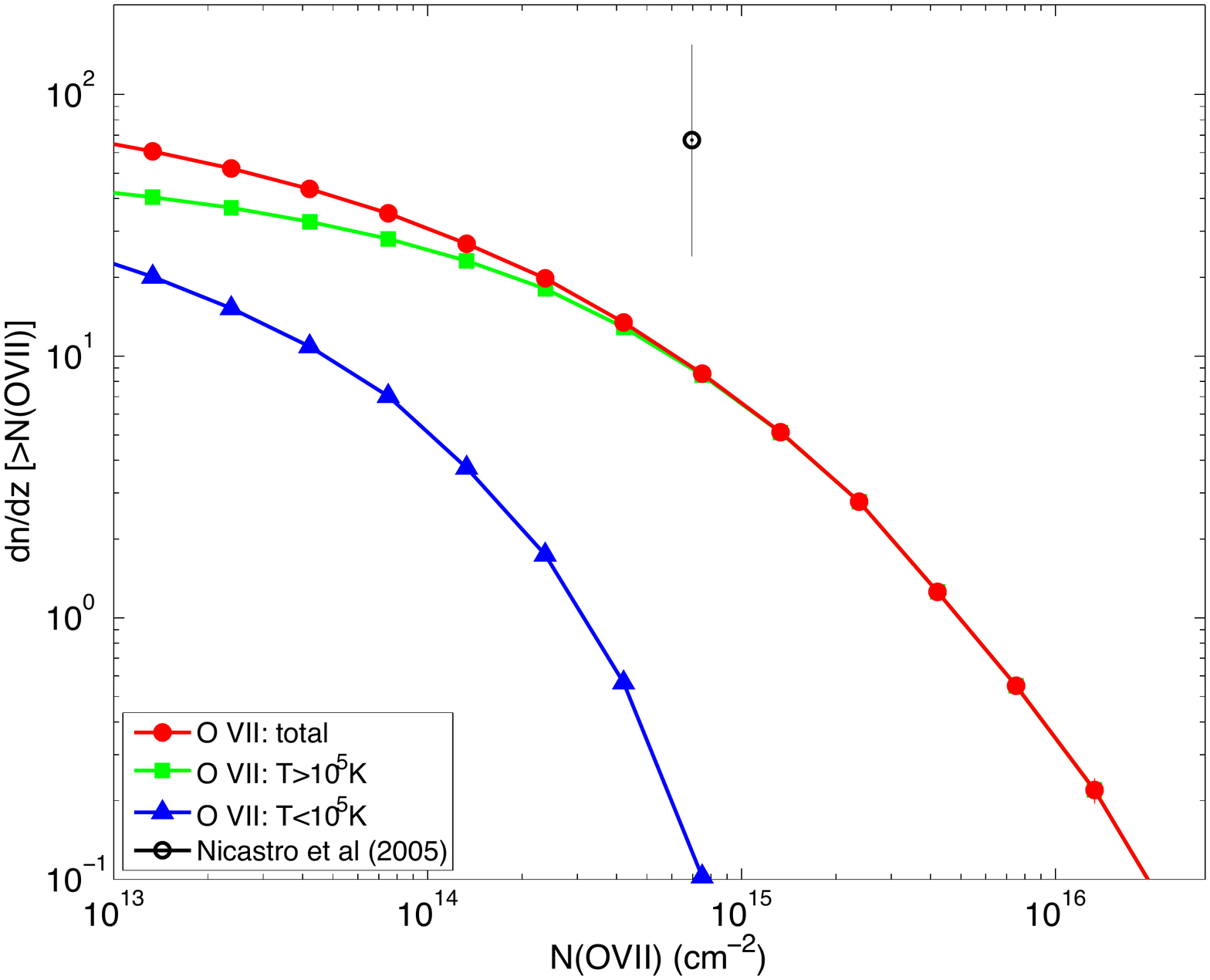}}
\vskip -0.4cm
\caption{
shows the cumulative O~VII line density as a function of column density,
defined to be the number of lines per unit redshift at the column density greater than the value at the x-axis.
The red solid dots, green squares and blue triangles
are the total, collisionally ionized and photo-ionized lines, respectively.
Also shown as a black open circle is the observation of \citet[][]{2005Nicastro} with $1\sigma$ errorbar.
Note that the quantify shown in the y-axis of  
Figure~\ref{fig:dndz_O6} is differential, not cumulative density.
}
\label{fig:dndz_O7}
\end{figure}

In \S 3.1 we show that some of the primary observable properties of simulated O~VI lines,
including line incidence rate, are in excellent agreement with observations.
In \S 3.2 we have shown various physical properties underlying the observables of both lines.
Before presenting quantitative coincidence rates between O~VI and O~VII lines,
it is useful to further check the O~VII line incidence rate to assess the self-consistency of our simulations
with extant observations.
Figure~\ref{fig:dndz_O7} shows the cumulative O~VII line density as a function of column density.
We also show the implied observed line density, under the assumption that the detection
reported by \citet[][]{2005Nicastro} is true.
We see that the claimed observational detection is about $2\sigma$ above or a factor of $\sim 7$ higher
than our predicted central value at the column density $\ge 7\times 10^{14}$cm$^{-2}$.
Our model is clearly in a more comfortable situation, if the claimed observational detection
turns out to be negative. 
As discussed in the introduction the detection reported by 
\citet[][]{2005Nicastro} is presently controversial.
This highlights the urgent need of higher sensitivity X-ray observations of this or other viable targets
that could potentially place strong constraints on the model.

\begin{figure}[H]
\centering
\resizebox{5.0in}{!}{\includegraphics[angle=0]{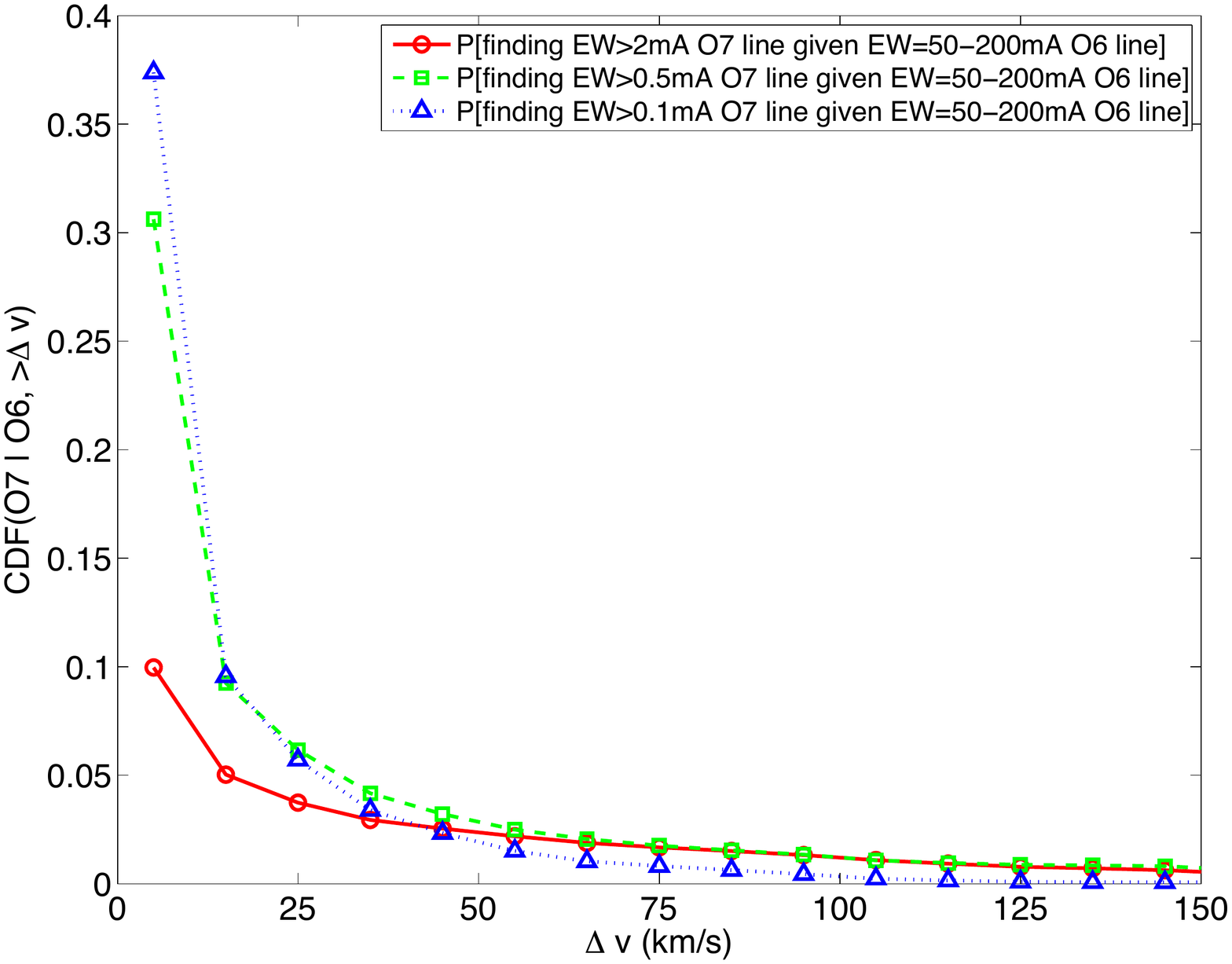}}
\vskip -1.0cm
\resizebox{5.0in}{!}{\includegraphics[angle=0]{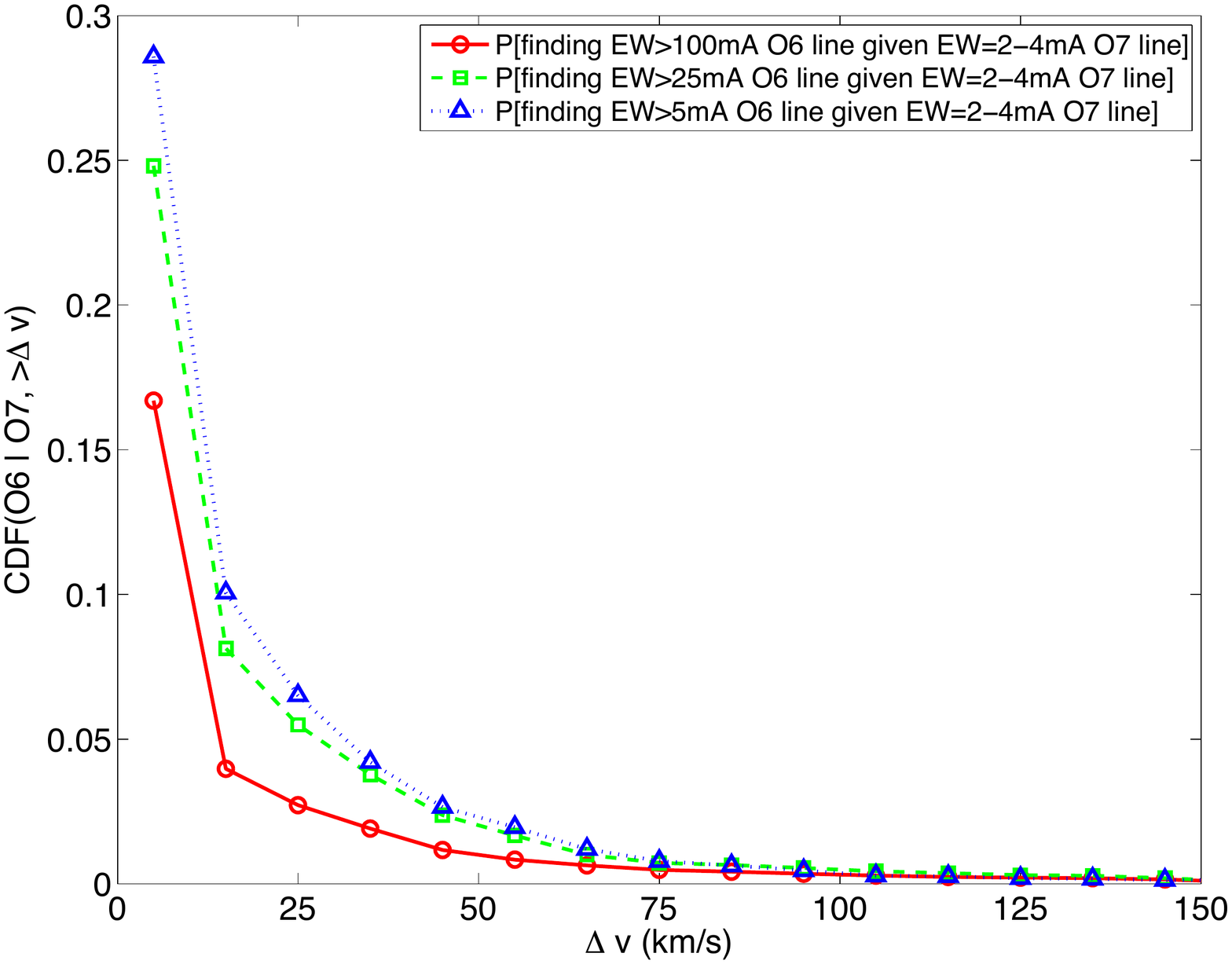}}
\vskip -0.4cm
\caption{
Top panel shows the cumulative probability distribution functions as a function of velocity displacement 
of having a coincidental O~VII line above the indicated equivalent width
for an O~VI line of a given equivalent width.
Bottom panel shows the cumulative probability distribution functions as a function of velocity displacement 
of having a coincidental O~VI line above the indicated equivalent width
for an O~VII line of a given equivalent width.
}
\label{fig:corr}
\end{figure}

We now turn to the coincidences between O~VI and O~VII lines.
The top panel of Figure~\ref{fig:corr} shows 
the cumulative probability distribution functions as a function of velocity displacement 
of having a coincidental O~VII line above the indicated equivalent width
for an O~VI line of a given equivalent width.
We see that O~VI lines of equivalent width in the range $50-200$mA
have $(38\%,31\%,10\%)$ probability of finding an O~VII line with equivalent
width greater than $(0.1,0.5,2)$mA within a velocity displacement of $150\kms$.
The vast majority of coincidental O~VII lines for O~VI lines for those equivalent widths in question
are concentrated within a velocity displacement of $\le 50\kms$ and more than 50\% at $\le 25\kms$.

The bottom panel of Figure~\ref{fig:corr} shows 
the cumulative probability distribution functions as a function of velocity displacement 
of having a coincidental O~VI line above the indicated equivalent width
for an O~VII line of a given equivalent width.
It is seen that for O~VII lines of equivalent width in the range $2-4$mA
have a $17-27\%$ probability of finding an O~VI line with equivalent
width in the range $5-100$mA within a velocity displacement of $150\kms$.
Likewise, the vast majority of coincidental O~VI lines for O~VII lines for those equivalent widths in question
are concentrated within a velocity displacement of $\le 50\kms$ and more than 80\% at $\le 25\kms$.

\begin{figure}[H]
\centering
\resizebox{5.0in}{!}{\includegraphics[angle=0]{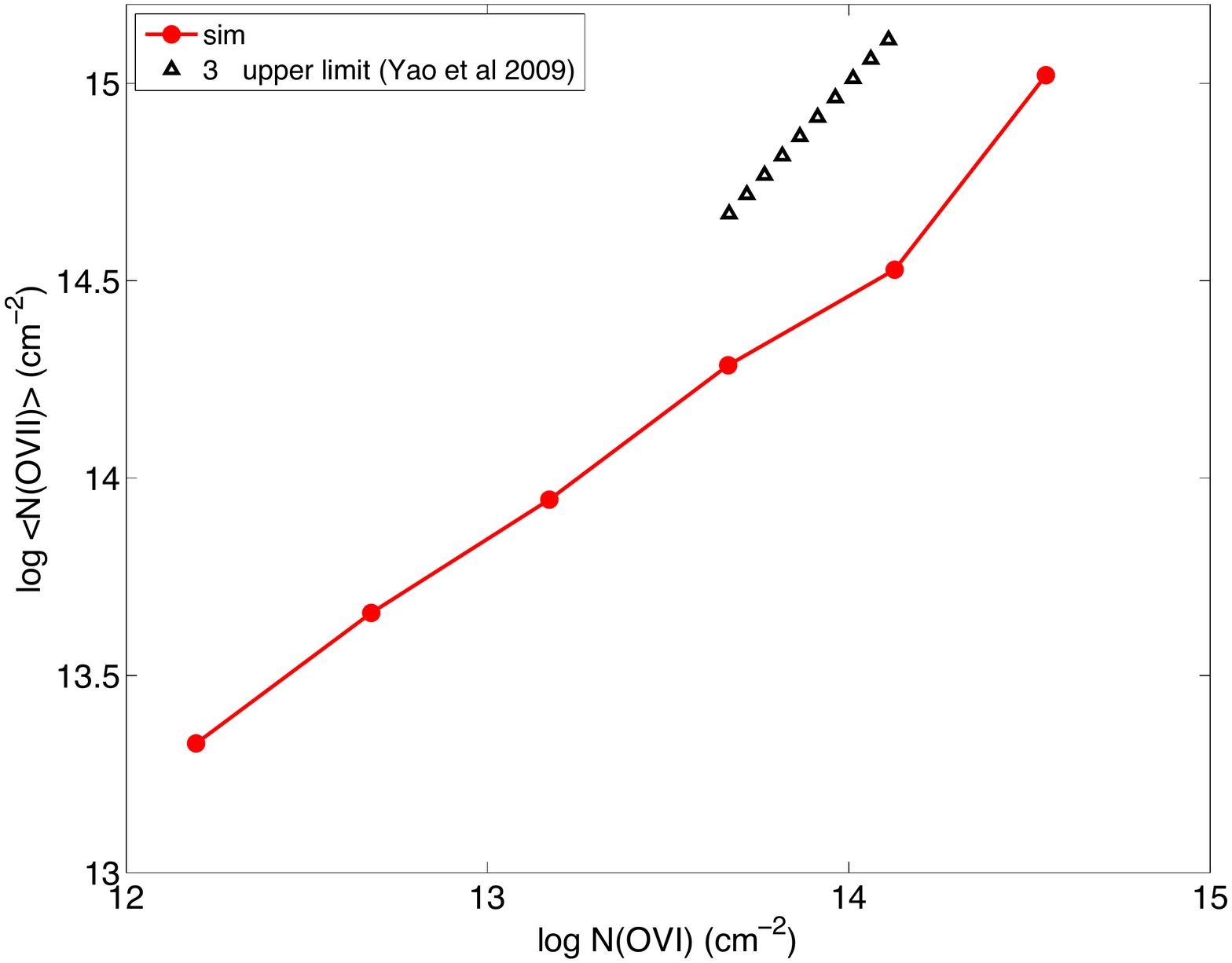}}
\vskip -1.0cm
\resizebox{5.0in}{!}{\includegraphics[angle=0]{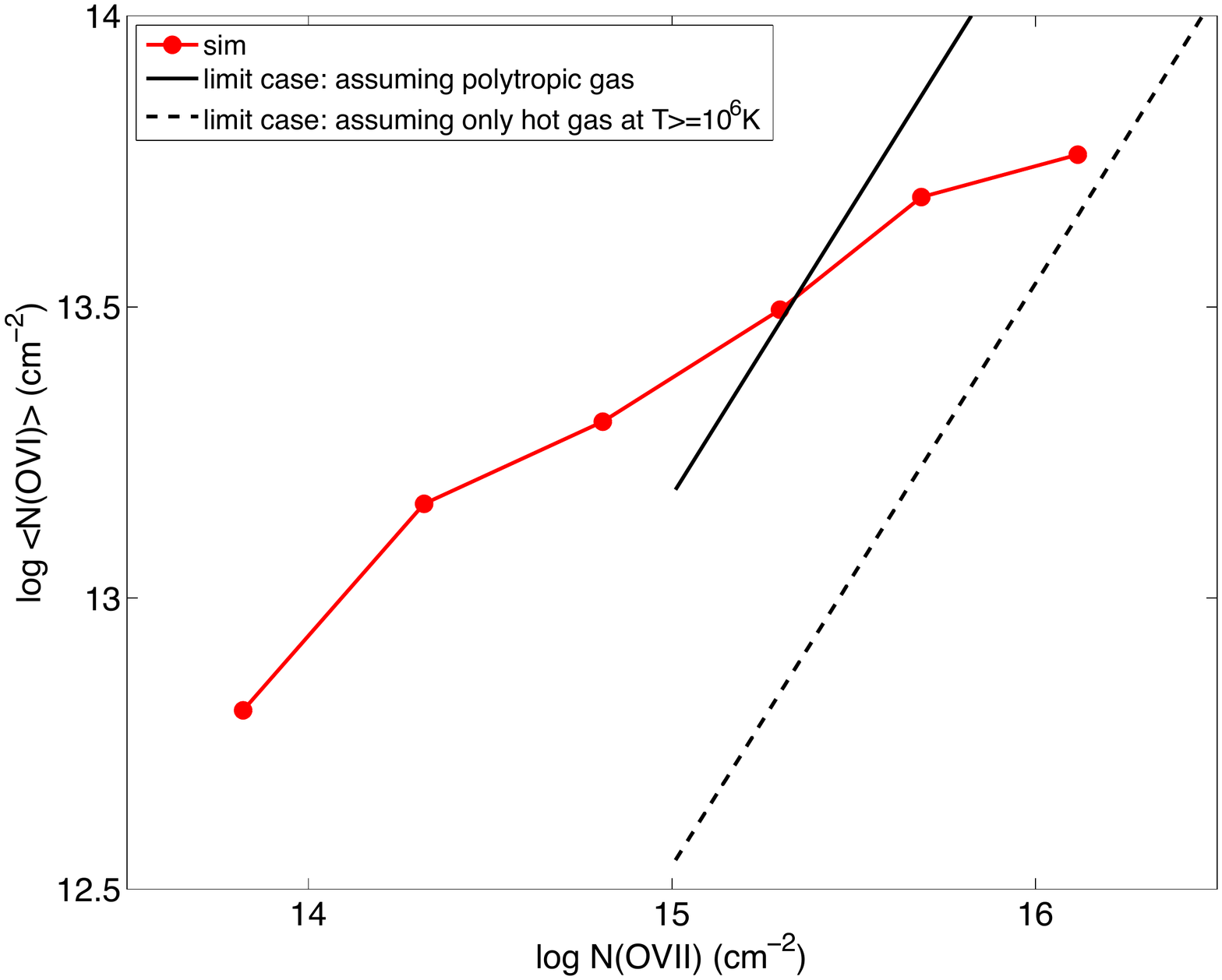}}
\vskip -0.4cm
\caption{
The top panel shows the expected mean O~VII column density 
at the location of detected O~VI lines of column density indicated by the x-axis.
Also shown as black triangles are $3\sigma$ upper limits from observations of \citet[][]{2009Yao}. 
The bottom panel shows the expected mean O~VI column density 
at the location of detected O~VII lines of column density indicated by the x-axis.
The solid and dashed straight lines are possible limit cases based on simple physical considerations.
}
\label{fig:N_O6O7}
\end{figure}

The results shown in Figure~\ref{fig:corr} presently can not be compared to observations,
because there is no definitive detection of O~VII absorbers,
although there are many detected O~VI absorbers.
Thus, we use the stacking method of \citet[][]{2009Yao} to enable a direct comparison with available observations.
The top panel of Figure~\ref{fig:N_O6O7} 
shows the expected mean O~VII column density 
at the location of detected O~VI lines of column density indicated by the x-axis,
compared to the $3\sigma$ upper limits from observations of \citet[][]{2009Yao} shown as black triangles.
We see that the non-detection of O~VII lines, or more precisely, a $3\sigma$ upper limit
on the mean column of O~VII lines for detected O~VI lines of column density in the range
$\log {\rm N(OVI)cm^2}=13.6-14.1$, is fully consistent with our simulations.
The reported $3\sigma$ upper limit is above the expected value by a factor of $2.5-4$.
This suggests that a factor of $\sim 10$ increase in sample size or sensitivity
will be able to yield a definitive detection of O~VII column density using the stacking technique
even without detection of individual O~VII absorbers.
The bottom panel of Figure~\ref{fig:N_O6O7} shows the expected mean O~VI column density 
at the location of detected O~VII lines of column density indicated by the x-axis.

It is evident from 
Figures~\ref{fig:corr},\ref{fig:N_O6O7} 
that O~VI and O~VII lines are coincidental only in a limited sense.
We attribute the limited coincidence of O~VII lines for O~VI lines 
primarily to two situations for O~VI producing regions.
A line of sight that intersects an O~VI producing region
does not necessarily intersect a strong O~VII producing region along the same line of sight,
either because the temperature of the overall region does not reach a high enough value to be strong O~VII bearing,
or because the intersected O~VI region is laterally an outskirt of an onion-like structure 
where the more central, higher temperature, O~VII region makes up 
a smaller cross section.
The former case should be ubiquitous, because weaker gravitational shocks that
produce regions of temperature, say, $10^{5.5}$K are more volume filling than  
stronger gravitational shocks giving rise to regions of temperature, say, $10^{6.0}$K. 
In addition, feedback shocks from star formation tend to be weaker than required
to collisionally produce O~VII at the spatial scales of interest here.
In other words, one expects to see many O~VI-bearing regions that have no associated O~VII-bearing sub-regions.
The latter case where hotter but smaller regions are surrounded by cooler regions
is expected to arise naturally around large virialized systems such as groups and clusters of galaxies.
A more quantitative but still 
intuitive physical check of the obtained results is not straight-forward,
without performing a much more detailed study of individual physical regions that produce O~VI and O~VII absorbers.
We shall reserve such a study for the future.

The situation of coincidental O~VI lines for given O~VII lines might appear to be less ambiguous at first sight 
in the sense that the hotter central, O~VII-producing regions should be surrounded by cooler regions and thus 
one might expect that the line of sight that intersects a strongly O~VII-producing region
should automatically intersect cooler regions that would show up as O~VI absorbers.
While it is true that a hot region is in general surrounded by cooler regions,
it is not necessarily true that a hot $10^6$K, O~VII-bearing gas is surrounded 
by significant $10^{5.5}$K, O~VI-bearing gas.
For example, one may have a post-shock region of temperature $10^6$K that is surrounded only by pre-shocked gas
that is much colder than $10^{5.5}$K.
We note that for a gas of $\delta=100$, $T=10^6$K and $Z=0.3\zsun$ at $z=0$,
its cooling time is $t_{cool}\sim 0.5t_H$ ($t_H$ is the Hubble time at $z=0$).
This means that O~VII-bearing WHIM gas at $\delta\le 100$, which has been heated up by shocks to $T\ge 10^6$K,
is unlikely to cool to $T\sim 10^{5.5}$K to become O~VI rich gas.
On the other hand, the cooling time for gas of $\delta=100$, ${\rm T}=10^{5.5}$K and $Z=0.1\zsun$ at $z=0$ is $\sim 0.05t_H$,
as noted earlier.
Thus, it is physically possible that sharp interfaces between hot (${\rm T}\ge 10^6$K) and
cold ${\rm T}\le 10^5$K gas develop.
The simulations do not include thermal conduction, which can be shown to be unimportant here.
The electron mean free path (mfp) is $0.44({\rm T}/10^6{\rm K})^2(\delta/100)^{-1}$kpc, adopting the standard Spitzer value.
The likely presence of magnetic fields (not treated here) would further reduce the mfp
by an order of magnitude \citep[e.g.,][]{1977Cowie}.
Thus, thermal conduction is insignificant and multi-phase media is expected to exist.

\begin{figure}[H]
\centering
\resizebox{5.0in}{!}{\includegraphics[angle=0]{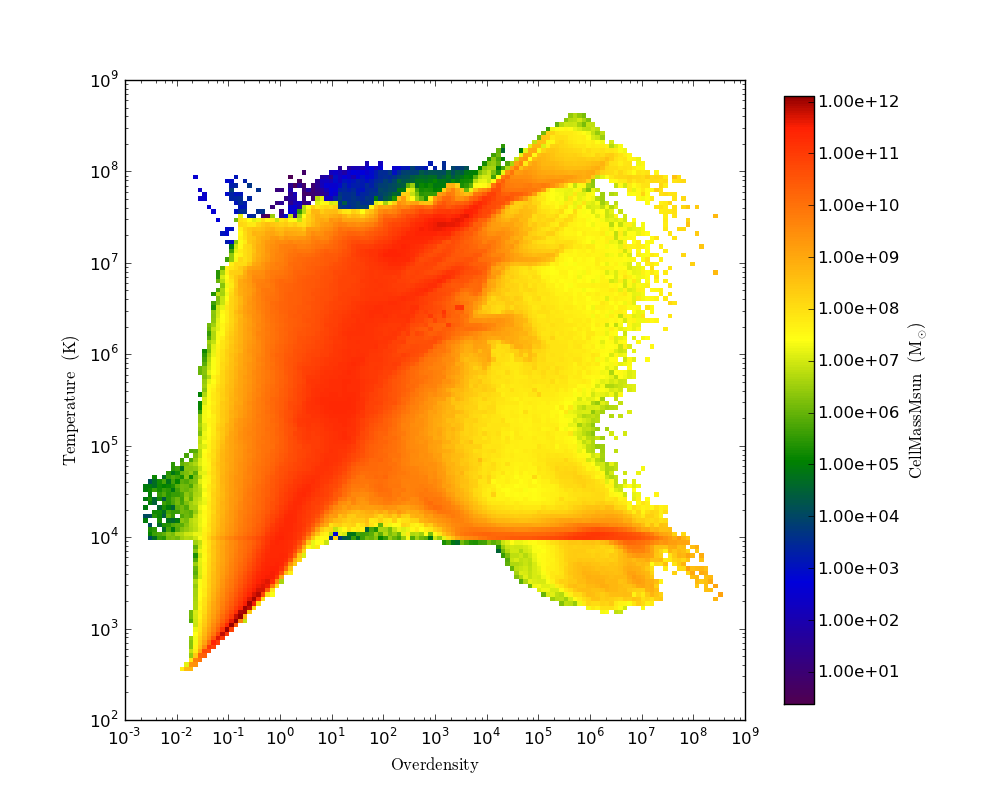}}
\vskip -0.0cm
\hskip -1.2cm
\resizebox{5.0in}{!}{\includegraphics[angle=0]{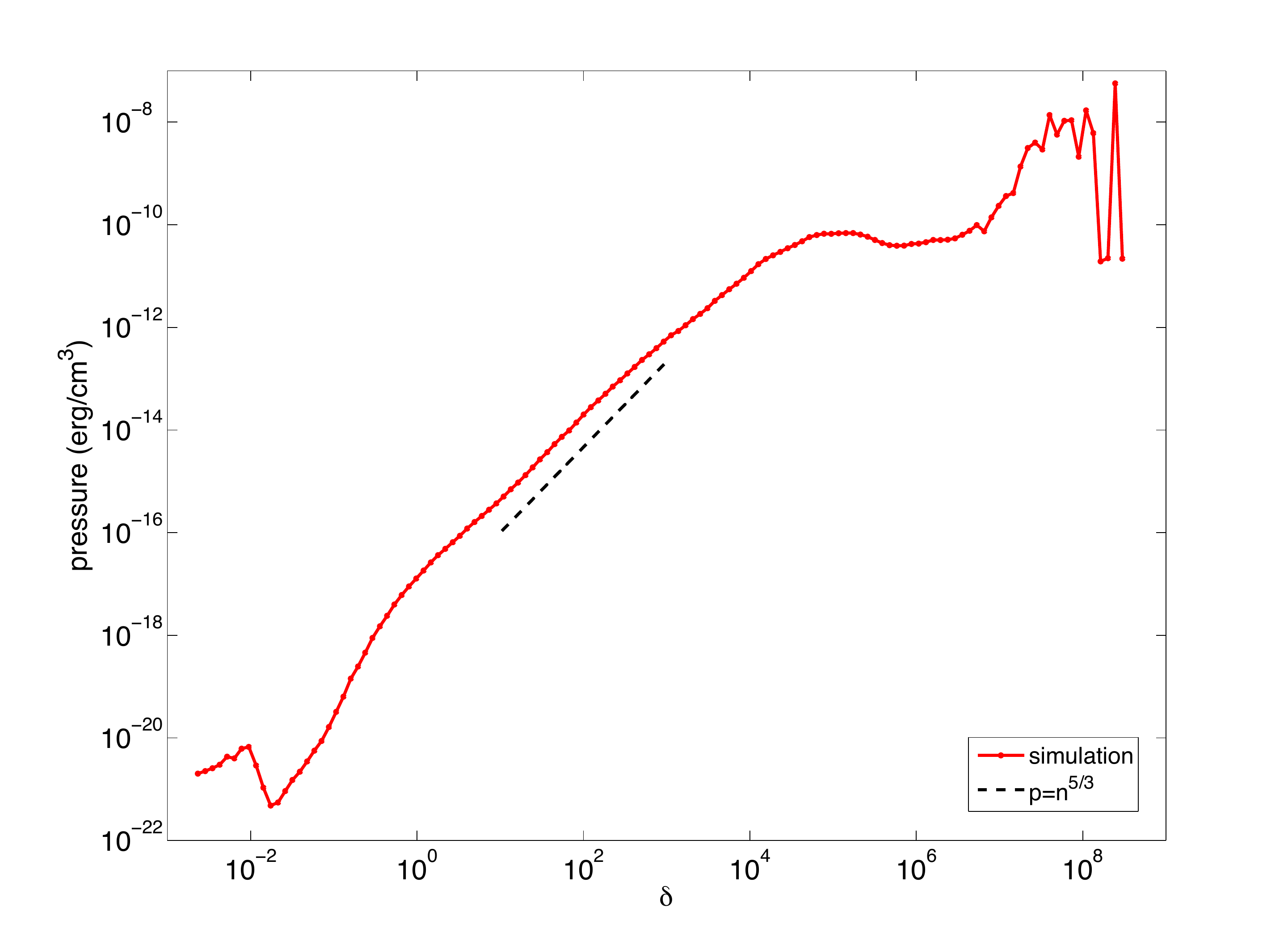}}
\vskip -0.4cm
\caption{
Top panel shows mass weighted phase diagram in the temperature-overdensity plane for the refined region in the C run.
The bottom panel shows the mean pressure as a function of overdensity averaged over all cells in the regions in the C run.
The black dashed line indicates the slope for polytropic gas of index $5/3$ (i.e., the adiabatic index),
which provides a good approximation to the simulation in the overdensity range $10-300$ that is
most pertinent to the absorbing WHIM in O~VI and O~VII.
}
\label{fig:phase}
\end{figure}

\begin{figure}[h!]
\hskip -0.7cm
\centering
\resizebox{5.0in}{!}{\includegraphics[angle=0]{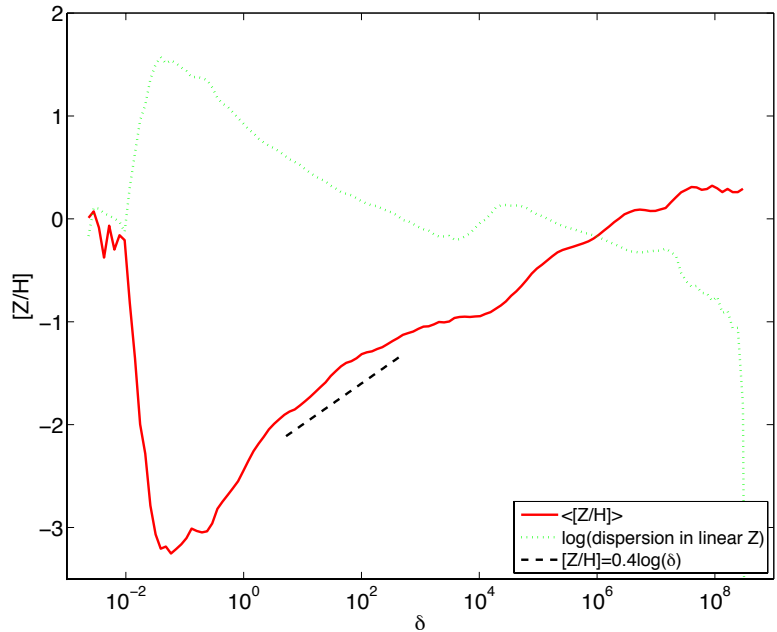}}
\hskip -0.5in
\vskip -0.4cm
\caption{
shows the mean gas metallicity (red solid curve) 
as a function of overdensity averaged over all cells in the refined region of the C run.
Also shown as the green dotted curve is the logarithm
of the dispersion in $Z$ (linear metallicity).
The black dashed line indicates the logarithmic slope of $0.4$,
which provides a good approximation to the simulation in the overdensity range $10-1000$ 
relevant to absorbing WHIM in O~VI and O~VII.
}
\label{fig:metal_overd}
\end{figure}

Equipped with this information and adopting the simpler implied geometry 
allows for a simple physical check of the results in the 
bottom panel of Figure~\ref{fig:N_O6O7}, as follows.
We will take two separate approaches to estimate this. The first approach assumes a polytropic gas 
in the temperature range relevant for O~VI and O~VII collisional ionization.
In the top panel of Figure~\ref{fig:phase} we show the entire 
temperature-overdensity phase diagram for the refined region in the C run.
Note that the gas density reaches about 1 billion times the mean gas density,
corresponding to $\sim 100$cm$^{-3}$, i.e., star formation regions.
For the regions of present relevance, the density range is about $10-300$ times the mean density,
illustrated by the upper part of the red tornado-like region near the middle of the plot.
It is useful to note that for this density range, the gas mass is dominated by 
gas in the temperature range of $10^5-10^7$K, i.e., WHIM.
Because of this reason, it is a valid exercise to compute the mean pressure as a function of 
overdensity, at least for the density range relevant for WHIM,
shown in the bottom panel of Figure~\ref{fig:phase}.
We see that for the WHIM overdensity range of $10-300$,
the adiabatic index $5/3$, shown as the dashed line,
provides an excellent approximation for the polytropic index of the gas.
It is also necessary to have a relation between gas metallicity as a function of gas overdensity,
shown in Figure~\ref{fig:metal_overd}.
We only note that the metallicity is 
generally an increasing 
function with density above one tenth of the mean density,
that the sharp rise of metallicity below one tenth of the mean density
is due to metal-enrich galactic winds escaping into the low density regions,
and that for our present purpose concerning the WHIM overdensity range of $10-300$ the metallicity roughly goes
as ${\rm Z}\propto \delta^0.4$, as indicated by the dashed line.

Given the information in Figures~\ref{fig:phase},~\ref{fig:metal_overd} we can now proceed to estimate
the expected O~VI column at a given O~VII column density, i.e., ${\rm<N(OVI)>/<N(OVII)>}$,
assuming both are dominated by collisional ionization.
The O~VI column density may be roughly approximated as 
${\rm <N(OVI)>\propto f(OVI) \Delta\log T(OVI) \rho(OVI) Z(OVI) L(OVI)}$,
where  ${\rm f(OVI)=0.22}$ at $\log{\rm T(OVI)/K=5.5}$, 
${\rm \Delta\log T(OVI)=0.2}$, ${\rm \rho(OVI)}$, ${\rm Z(OVI)}$ and ${\rm L(OVI)}$
are the peak collisional ionization fraction for O~VI, 
the FWHM of the logarithmic temperature of the collisional ionization peak (see 
the blue curve in the top panel of Figures~\ref{fig:This_O6}),
the density of the O~VI absorbing gas,
the metallicity of the O~VI absorbing gas
and the physical thickness of the O~VI absorbing gas, respectively.
We have an exactly analagous relation for O~VII, with ${\rm f(OVII)=1}$, $\log{\rm T(OVII)=6}$, ${\rm \Delta\log T(OVII)=0.7}$.
With an additional assumption that the characteristic thickness 
at a given density goes as 
${\rm L(OVI)\propto\rho(OVI)^{1/3}}$ (i.e., mass distribution across log density is roughly uniform),
we can now evaluate the column density ratio 
\begin{eqnarray}
\label{eq:ratio}
{\rm <N(OVI)>\over <N(OVII)>} &=& 
 {\rm f(OVI)\over f(OVII)}
 {\rm \Delta\log T(OVI)\over \Delta\log T(OVII)}
 {\rm \rho(OVI)\over \rho(OVII)} 
 {\rm Z(OVI)\over Z(OVII)} 
 {\rm L(OVI)\over L(OVII)} \nonumber \\
 &=& 
 {\rm f(OVI)\over f(OVII)}
 {\rm \Delta\log T(OVI)\over \Delta\log T(OVII)}
\left({\rm T(OVI)\over T(OVII)}\right)^{(2/3+\alpha)/(\gamma-1)} = 0.015,
\end{eqnarray}

\noindent
where $\alpha=0.4$ and $\gamma=5/3$ are used, as indicated in 
Figures~\ref{fig:metal_overd} and Figures~\ref{fig:phase}, respectively.
This resulting ratio is shown as the solid line in the bottom panel of Figure~\ref{fig:N_O6O7},
which we expect to an approximate upper limit of the true ratio, since it implies the presence 
of O~VI-bearing gas for every O~VII line.

Our second approach likely gives an approximate lower bound on ${\rm <N(OVI)>/<N(OVII)>}$.
We assume that the O~VII-bearing gas is at the peak temperature of $10^6$K and is 
surrounded by gas that has a temperature that is much lower than $10^{5.5}$K (neglecting
photoionization for the moment),
in which case the coincidental O~VI line is produced by the same temperature gas that
produces the O~VII line, giving
\begin{eqnarray}
\label{eq:ratio}
{\rm <N(OVI)>\over <N(OVII)>} &=& {\rm f(OVI) (T=10^6K)\over f(OVII) (T=10^6K)} \nonumber \\
 &=& 0.0035,
\end{eqnarray}
\noindent
which is shown as the dashed line in the bottom panel of Figure~\ref{fig:N_O6O7}.
Admittedly, our approaches to estimate the column density ratios
are quite simplistic.
Nevertheless, we think they capture some of the essential underlying relationships between O~VI-bearing gas
and O~VII-bearing gas in the collisional ionization dominated regime
and it is reassuring that they are consistent with detailed calculations.
Note that at ${\rm N(OVII)<10^{15}}$cm$^{-2}$ photoionization becomes important,
especially for related O~VI lines, hence our simple physical illustration breaks down 
in that regime.

\section{Conclusions}

Utilizing high resolution ($0.46h^{-1}$kpc), adaptive mesh-refinement Eulerian cosmological hydrodynamic simulations 
we examine properties of O~VI and O~VII absorbers in the warm-hot intergalactic medium (WHIM) at $z=0$,
along with a physical examination.
We find that our new high resolution simulations are in broad agreement with all other simulations
with respect to the thermal distribution of baryons in the present universe.
In particular, we find that about $40\%$ of the intergalactic medium is in the WHIM.
We find that our simulations are in excellent agreement with observed properties of O~VI absorbers,
including line incidence rate, Doppler width-column density relation, and consistent with 
observed Doppler width-temperature relation. 
Physical properties of O~VI and O~VII absorbers are given, including inter-relations between
metallicity, temperature, density, Doppler width, to facilitate a coherent understanding.
We highlight some of the important or new findings.

(1) We find that strong O~VI absorbers are predominantly collisionally ionized,
whereas for weaker absorbers the contributions from  photoionization become progressively more important.
We find that $(39\%, 43\%, 61\%)$ of O~VI absorbers in the 
column density ranges of $\log{\rm N(OVI)cm^2}=(12.5-13,13-14,>14)$ have temperature greater than $10^5$K.
This may be contrasted with the results of
\citet[][]{2009Oppenheimer} where low temperature ($\sim 10^4$K), high metallicity, photoionized O~VI absorbers
dominate even at high column densities ($\log{\rm N(OVI)cm^2}>14$).
We concur that lack of metal mixing in SPH simulations, which in turn 
causes overcooling of high-metallicity feedback SPH particles, most severely in high density regions, 
may be able to account for the discrepancy.
We suggest that cross correlations between strong [${\rm N(OVI)}\ge 10^{14}$cm$^{-2}$] O~VI absorbers and galaxies 
on $\sim 100$kpc scales may be able to differentiate between the models.

(2) Velocity structures within absorbing regions are 
a significant, and for large Doppler width clouds, a dominant contributor to the Doppler widths
of both O~VI and O~VII absorbers.
Doppler width is thus a poor indicator of temperature.

(3) Quantitative prediction is made for the presence of broad and shallow O~VI lines,
which current observations have largely failed to detect.
Upcoming observations by COS may be able to provide a test.

(4) The coincidence rates between O~VI and O~VII lines are found to be small, for which physical explanations are given.
We find that the reported $3\sigma$ upper limit on the mean column density of coincidental O~VII lines
at the location of detected O~VI lines by \citet[][]{2009Yao} 
is above the predicted value by a factor of $2.5-4$,
implying that a factor of $\sim 10$ increase in sample size or sensitivity
will be able to yield a definitive detection of O~VII column density using the stacking technique
even without detection of individual O~VII absorbers.

(5) We show that, if the previously claimed observational detection of O~VII lines by \citet[][]{2005Nicastro} is true,
our predicted O~VII line density is $2\sigma$ below that.
This shows that higher sensitivity X-ray observations of this or other viable targets
will be very useful to potentially place strong constraints on the model.

\vskip 1cm

I would like to thank Dr. M.K.R. Joung for help on
generating initial conditions for the simulations and running a portion
of the simulations and Greg Bryan for help with Enzo code.
I would like to thank the referee Mike Shull for critical and constructive reports.
I would like to thank Dr. Edward Jenkins for a careful reading of the manuscript
and helpful discussion,
Dr. Charles Danforth for kindly providing the observational data and useful discussion,
Dr. Jeremiah P. Ostriker for useful discussion and Drs. John Wise, Matthew Turk and Cameron Hummels
for help with visualization program yt \citep[][]{2011Turk}.
Computing resources were in part provided by the NASA High-
End Computing (HEC) Program through the NASA Advanced
Supercomputing (NAS) Division at Ames Research Center.
This work is supported in part by grants NNX08AH31G and NAS8-03060. 
The simulation data are available from the author upon request.

\bibliographystyle{apj}
\bibliography{astro}

\end{document}